\documentclass[12pt,twoside]{article}
\usepackage[mathscr]{eucal}
\usepackage{amsmath,amsfonts,amssymb,amsthm}
\usepackage[matrix,arrow]{xy}
\usepackage{mathabx}
\usepackage{amsthm,amsmath,latexsym,amssymb,amsfonts,amsbsy}
\usepackage{wrapfig}
\usepackage{graphicx}
\usepackage{float}

\usepackage[breaklinks]{hyperref}
\usepackage{amsmath}
\usepackage{amssymb}
\usepackage{braket}
\usepackage[T1]{fontenc}

\usepackage{tikz,pgf}
\usetikzlibrary{shapes}
\usetikzlibrary{calc}
\usetikzlibrary{decorations.pathmorphing}
\usetikzlibrary{decorations.pathreplacing,shapes.misc}
\usetikzlibrary{positioning}
\usetikzlibrary{arrows}
\usetikzlibrary{decorations.markings}

\def\be{\begin{equation}}
\def\ee{\end{equation}}
\def\ba{\begin{array}}
\def\ea{\end{array}}

\def\dps{\displaystyle}
\newcommand{\half}{\frac{1}{2}}

\def\a{\tilde{1}}
\def\b{\tilde{2}}
\def\1{\tilde{1}}
\def\2{\tilde{2}}
\def\3{\tilde{3}}

\newdimen\tableauside\tableauside=1.0ex
\newdimen\tableaurule\tableaurule=0.4pt
\newdimen\tableaustep
\def\phantomhrule#1{\hbox{\vbox to0pt{\hrule height\tableaurule
width#1\vss}}}
\def\phantomvrule#1{\vbox{\hbox to0pt{\vrule width\tableaurule
height#1\hss}}}
\def\sqr{\vbox{%
  \phantomhrule\tableaustep

\hbox{\phantomvrule\tableaustep\kern\tableaustep\phantomvrule\tableaustep}%
  \hbox{\vbox{\phantomhrule\tableauside}\kern-\tableaurule}}}
\def\squares#1{\hbox{\count0=#1\noindent\loop\sqr
  \advance\count0 by-1 \ifnum\count0>0\repeat}}
\def\tableau#1{\vcenter{\offinterlineskip
  \tableaustep=\tableauside\advance\tableaustep by-\tableaurule
  \kern\normallineskip\hbox
    {\kern\normallineskip\vbox
      {\gettableau#1 0 }%
     \kern\normallineskip\kern\tableaurule}%
  \kern\normallineskip\kern\tableaurule}}
\def\gettableau#1 {\ifnum#1=0\let\next=\null\else
  \squares{#1}\let\next=\gettableau\fi\next}

\renewcommand{\tilde}{\widetilde}

\newcommand{\bref}[1]{\textbf{\ref{#1}}}

\def\cF{\mathcal{F}}

\def\cO{\mathcal{O}}

\numberwithin{equation}{section} \makeatletter
\@addtoreset{equation}{section}

\def\be{\begin{equation}}
\def\ee{\end{equation}}
\def\ba{\begin{array}}
\def\ea{\end{array}}

\def\dps{\displaystyle}

\def\ba{\begin{array}}
\def\ea{\end{array}}

\def\dps{\displaystyle}

\begin{document}
\begin{flushright}
FIAN-TD-2015-04 \\
\end{flushright}

\vspace{1cm}

\begin{center}

{\Large\textbf{Classical conformal blocks via  AdS/CFT correspondence}}

\vspace{.9cm}

{\large Konstantin  Alkalaev$^{\;a,b}$ and   Vladimir Belavin$^{\;a,c}$}

\vspace{0.5cm}

\textit{$^{a}$I.E. Tamm Department of Theoretical Physics, \\P.N. Lebedev Physical
Institute,\\ Leninsky ave. 53, 119991 Moscow, Russia}

\vspace{0.5cm}

\textit{$^{b}$Moscow Institute of Physics and Technology, \\
Dolgoprudnyi, 141700 Moscow region, Russia}

\vspace{0.5cm}

\textit{$^{c}$ Department of Quantum Physics, \\ 
Institute for Information Transmission Problems, \\
Bolshoy Karetny per. 19, 127994 Moscow, Russia}

\vspace{0.5cm}

\thispagestyle{empty}


\end{center}

\begin{abstract}
We continue to develop the holographic  interpretation of classical conformal blocks in terms of particles propagating in an asymptotically $AdS_3$ geometry. We study $n$-point block with two heavy and $n-2$ light fields. Using the worldline approach we propose and explicitly describe the corresponding bulk configuration, which consists of $n-3$ particles propagating in the conical defect background produced by the heavy fields. We test this general picture in the case of five points. Using the special combinatorial representation of the Virasoro conformal block we compute $5$-point classical block and find the exact correspondence with the bulk worldline action. In particular, the bulk analysis relies upon the special perturbative procedure which treats the $5$-point case as a deformation of the $4$-pt case.

\end{abstract}

\section{Introduction}

Two-dimensional conformal field theories with Virasoro algebra have dual description in terms of pure gravity in three-dimensional  asymptotically anti-de Sitter spacetime \cite{Brown:1986nw}. The correspondence 
establishes a connection between various important objects  on two sides of the AdS/CFT correspondence. 
While for finite values of the central charge $c$ this connection is still not enough studied, in the case where the central charge tends  to infinity $c\rightarrow\infty$ there are many interesting results found recently. Bulk computations in this limit  are performed using a saddle point
approximation  and allow to reconstruct the content of AdS  theory  from the  boundary CFT configuration
(see, \textit{e.g.}, \cite{Heemskerk:2009pn,ElShowk:2011ag,Fitzpatrick:2012cg,Jackson:2014nla,deBoer:2014sna} and references therein). 
Remarkably, in this case holographic computation can be performed  not only for correlation functions 
(representing, of course, our main interest) but also for more fundamental objects -- conformal block  functions   \cite{Belavin:1984vu}.

\begin{figure}[H]
\centering
\begin{tikzpicture}
\draw [line width=1pt] (30,0) -- (32,0);
\draw [line width=1pt] (32,0) -- (32,2);
\draw [smooth, tension=1.0, line width=1pt, decorate, decoration = {snake, segment length = 2mm, amplitude=0.4mm}] (32,0) -- (34,0);
\draw [line width=1pt] (34,0) -- (34,2);
\draw [smooth, tension=1.0, line width=1pt, decorate, decoration = {snake, segment length = 2mm, amplitude=0.4mm}] (34,0) -- (36,0);
\draw [line width=1pt] (36,0) -- (36,2);
\draw [smooth, tension=1.0, line width=1pt, decorate, decoration = {snake, segment length = 2mm, amplitude=0.4mm}] (36,0) -- (38,0);
\draw [line width=1pt] (38,0) -- (38,2);
\draw [smooth, tension=1.0, line width=1pt, decorate, decoration = {snake, segment length = 2mm, amplitude=0.4mm}] (38,0) -- (40,0);
\draw [line width=3pt] (40,0) -- (40,2);
\draw [line width=3pt] (40,0) -- (42,0);


\draw (29,-0) node {$z_1, \Delta_1$};
\draw (32,2.5) node {$z_2, \Delta_{2}$};
\draw (38,2.5) node {$z_{n-2}, \Delta_{n-2}$};

\draw (35,2.5) node {$\cdots\cdots$};
\draw (43,-0) node {$z_n, \Delta_n$};
\draw (41.3,1.8) node {$z_{n-1}, \Delta_{n-1}$};

\draw (33,-0.6) node {$\tilde\Delta_1$};
\draw (39,-0.6) node {$\tilde\Delta_{n-3}$};
\draw (37,-0.6) node {$\tilde\Delta_{n-2}$};
\draw (35,-0.6) node {$\cdots\cdots$};


\fill (32,0) circle (0.8mm);
\fill (30,0) circle (0.8mm);
\fill (32,2) circle (0.8mm);

\fill (34,0) circle (0.8mm);
\fill (34,2) circle (0.8mm);

\fill (36,0) circle (0.8mm);
\fill (36,2) circle (0.8mm);

\fill (38,0) circle (0.8mm);
\fill (38,2) circle (0.8mm);

\fill (40,0) circle (0.8mm);


\end{tikzpicture}
\caption{The $n$-point conformal block.  In the case of the heavy-light conformal block two bold lines on the right are heavy fields, while external fields and intermediate fields depicted respectively by solid lines and wavy lines are light. Using the projective invariance one can fix the coordinates of three fields as $z_1=0, z_{n-1}=1,z_n=\infty$.}
\label{block}
\end{figure}
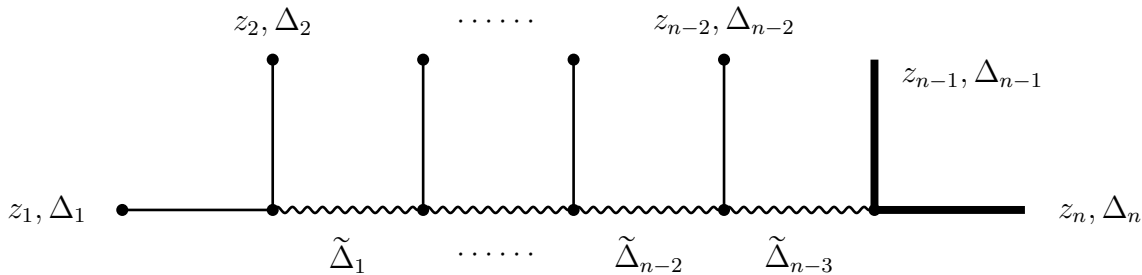

In the regime $c\rightarrow\infty$ we are dealing with the (semi)classical conformal blocks.
Recently, the construction of the classical conformal blocks in the context of the AdS/CFT correspondence  was investigated \cite{Fitzpatrick:2014vua,Asplund:2014coa,Hijano:2015rla}.
There was established a connection between the classical four-point block on the sphere with two heavy and two light operators   and the classical action of some worldlines combination in asymptotically $AdS_3$ geometry.
Conformal blocks represent building blocks in the construction of the correlation functions. They are  holomorphic functions of the coordinates $z_i$ of fields conveniently defined using the dual {\it pant decomposition} diagram.
In Fig. \bref{block} the dual diagram for $n$-point block is given (the meaning of  two bold lines is explained later). It represents the contribution from the states in the  Virasoro 
representations with highest  weights $\tilde \Delta_1,...,\tilde \Delta_{n-3}$ (internal lines) to the correlation function of the  fields  with conformal dimensions $\Delta_i$  (external lines).
We denote the corresponding  $n$-point 
conformal block depending on the parameters of the external and internal conformal dimensions (as well as on the central charge)
\be
\mathcal{F}(z_1,...,z_n|\Delta_1,...,\Delta_n;\tilde{\Delta}_1,...,\tilde{\Delta}_{n-3};c)\;.
\ee
The antiholomorphic block is defined similarly, replacing holomorphic coordinates $z_k$ by antiholomorphic $\bar{z}_k$. The
construction of the correlation functions involves the  summation  of the products of the  holomorphic and antiholomorphic blocks  over all possible intermediate channels 
$\tilde \Delta_i$ weighted with the structure constants of the operator algebra.

There exist many evidences ( see, \textit{e.g.}, \cite{Zamolodchikov1986,Harlow:2011ny}) that  in the classical limit the conformal blocks must  exponentiate as
\be
\label{classblockdef}
\lim_{c\rightarrow\infty}\mathcal{F}(z_1,...,z_n|\Delta_1,...,\Delta_n;\tilde{\Delta}_1,...,\tilde{\Delta}_{n-3};c)\sim \exp\big\{c f(z_1,...,z_n|\epsilon_1,...,\epsilon_n;\tilde{\epsilon}_1,...,\tilde{\epsilon}_{n-3})\big\}\;,
\ee
where $\epsilon_k=\frac{\Delta_k}{c}$ and $\tilde{\epsilon_k}=\frac{\tilde{\Delta}_k}{c}$  are called {\it classical dimensions}
and $f(z_1,...,z_n|\epsilon_1,...,\epsilon_n;\tilde{\epsilon}_1,...,\tilde{\epsilon}_{n-3})$ is the classical conformal block representing our main interest. 

There are different possible classical limits of the conformal blocks dependent on the behaviour of the
classical dimensions $\epsilon_i$ and $\tilde\epsilon_i$ \cite{Zamolodchikov:1995aa,Fitzpatrick:2014vua,Hijano:2015rla,Fitzpatrick:2015zha}.
If  the classical dimension remains finite in the classical limit, the corresponding field is called
"heavy" otherwise it is "light".  If all fields are light we are dealing with the \textit{global} $sl(2)$ conformal block while
in the opposite case where all fields are heavy we are dealing with the {\it proper} classical block. 
All other possibilities that could be referred to as heavy-light classical blocks can be considered as an interpolation between these two extreme regimes.

In this paper we study $n$-point classical conformal block in the context of the AdS/CFT correspondence. We are interested in the case where the classical conformal dimension of two fields $\epsilon_{n-1}$ and $\epsilon_n$ are heavy.
This fact is expressed using bold lines in Fig. \bref{block}.  The heavy operators with equal conformal dimensions $\epsilon_{n} = \epsilon_{n-1} \equiv \epsilon_h$ produce an asymptotically $AdS_3$ geometry identified either with a deficit angle or BTZ black hole geometry. 
To describe the interpolation between  the proper conformal block  and  the conformal block with only two heavy fields
it is instructive to introduce a scale factor $\delta$ \cite{Hijano:2015rla} that we call \textit{a lightness parameter}.
Schematically, provided that all except two dimensions are rescaled as   $\epsilon \rightarrow \delta \epsilon$ and 
$\tilde\epsilon \rightarrow \delta \tilde\epsilon$ there appear  a series expansion  
\be
\label{lightness}
f(z|\epsilon, \tilde\epsilon)=f_{\delta}(z|\epsilon, \tilde\epsilon)\, \delta +f_{\delta^2}(z|\epsilon, \tilde\epsilon) \, \delta^2\,+\,...\;.
\ee
The leading contribution $f_{\delta}(z)$ yields the conformal block with only  two heavy fields, while taking into account sub-leading contributions approximate the proper conformal block on the left hand side. 

From the AdS/CFT perspective, the boundary  fields are realized via particular graph of worldlines of $n-3$ classical point probes  propagating in the background geometry formed by the two  boundary heavy fields. On the bulk side, the heavy field dimensions are expressed via the mass parameter $\alpha^2 = 1 - 4 \epsilon_h$ of the background metric ($\alpha^2 >0$ for a conical defect, $\alpha^2<0$ for the BTZ black hole), while the light dimensions are identified with AdS masses via the standard formula for scalar fields, $m^2 = \Delta(\Delta-4)/R_{AdS}$.
Note that in this case the lightness parameter $\delta$ introduced in \eqref{lightness} measures a backreaction of the background on a probe.

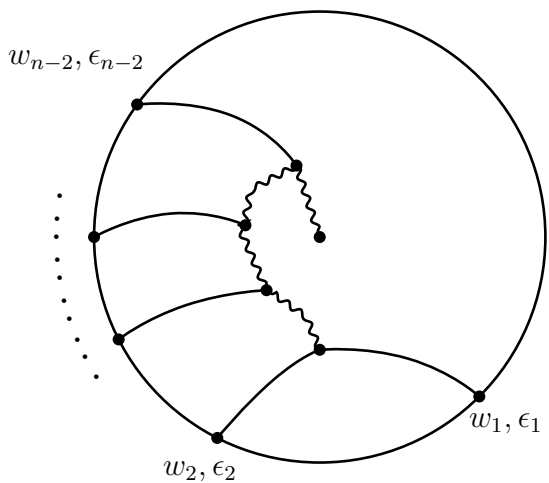
\begin{figure}[H]
\centering
\begin{tikzpicture}[line width=1pt]
\draw (0,0) circle (3cm);

\foreach \a in {1,2,...,40}{
\draw (\a*360/40: 3.5cm) coordinate(N\a){};
\draw (\a*360/40+5: 3.5cm) coordinate(D\a){};

\draw (\a*360/40: 4cm) coordinate(A\a){};
\draw (\a*360/40: 3cm) coordinate(K\a){};
\draw (\a*360/40: 2.5cm) coordinate(F\a){};
\draw (\a*360/40: 2cm) coordinate(L\a){};
\draw (\a*360/40: 1.5cm) coordinate(I\a){};
\draw (\a*360/40: 1.1cm) coordinate(J\a){};
\draw (\a*360/40: 1cm) coordinate(M\a){};
\draw (\a*360/40: 0.8cm) coordinate(C\a){};
;}


\draw plot [smooth, tension=1.0, line width=1pt] coordinates {(K35) (L34) (I30)};
\draw plot [smooth, tension=1.0, line width=1pt] coordinates {(K27) (L28) (I30)};
\draw plot [smooth, tension=1.0, line width=1pt] coordinates {(K23) (L23) (M25)};
\draw plot [smooth, tension=1.0, line width=1pt] coordinates {(K20) (L19) (M19)};
\draw plot [smooth, tension=1.0, line width=1pt] coordinates {(K16) (L14) (M12)};


\draw [smooth, tension=1.0, line width=1pt, decorate, decoration = {snake, segment length = 2mm, amplitude=0.4mm}] (M12)  -- (0,0);

\draw  [smooth, tension=1.0, line width=1pt, decorate, decoration = {snake, segment length = 2mm, amplitude=0.4mm}]  (I30) -- (J29) -- (M25);

\draw  [smooth, tension=1.0, line width=1pt, decorate, decoration = {snake, segment length = 2mm, amplitude=0.4mm}]  (M12) -- (J16) -- (M19);

\draw  [smooth, tension=1.0, line width=1pt, decorate, decoration = {snake, segment length = 2mm, amplitude=0.4mm}]  (J19) -- (M25);

\fill (I30) circle (0.8mm);
\fill (M25) circle (0.8mm);
\fill (K35) circle (0.8mm);
\fill (K27) circle (0.8mm);
\fill (K23) circle (0.8mm);
\fill (0,0) circle (0.8mm);
\fill (M19) circle (0.8mm);
\fill (M12) circle (0.8mm);
\fill (K20) circle (0.8mm);
\fill (K16) circle (0.8mm);



\draw (N27) node {$w_2,  \epsilon_2$};
\draw (N35) node {$w_1,  \epsilon_1$};
\draw (A16) node {$w_{n-2},  \epsilon_{n-2}$};

\draw (D19) node {{\bf.}};
\draw (N19) node {{\bf.}};

\draw (D20) node {{\bf.}};
\draw (N20) node {{\bf.}};

\draw (D21) node {{\bf.}};
\draw (N21) node {{\bf.}};

\draw (D22) node {{\bf.}};
\draw (N22) node {{\bf.}};

\draw (D23) node {{\bf.}};
\draw (N23) node {{\bf.}};

\end{tikzpicture}
\caption{Multi-particle graph embedded into a constant time slice of a conical defect geometry. Solid lines represent external particles, wavy lines represent intermediate particles. The original heavy fields produce the background geometry with the singularity placed in the center  representing a cubic vertex of two heavy fields  and a light intermediate field.  } 
\label{bulk}
\end{figure}

Using symmetry arguments  three-dimensional bulk analysis can be consistently reduced to a constant time slice identified with  a two-dimensional disk. 
The corresponding bulk configuration of the worldlines is shown in Fig. \bref{bulk}. To draw the bulk worldline graph in Fig. \bref{bulk} associated to a given boundary diagram  in Fig. \bref{block} we have a simple mnemonic rule: the boundary diagram is to be  pasted into the disc in  such a way  that outer ends of solid lines are attached to distinguished points on the boundary circle, while bold lines are collapsed into the origin of coordinates with one intermediate wavy line attached.

It was argued in \cite{Fitzpatrick:2014vua,Asplund:2014coa,Hijano:2015rla} that the classical conformal block coincides with the bulk classical action  
\be
\label{block-action}
S_{cl}^{bulk} = z^{\gamma}f_{\delta}(z|\epsilon, \tilde \epsilon)\;,
\ee
where 
\be
S_{cl}^{bulk}=\sum_{i=1}^{n-2} \epsilon_i\, L_i+\sum_{i=1}^{n-3} \tilde{\epsilon}_i\, \tilde L_i\;,
\ee
and $L_i$ and $\tilde L_i$  are lengths of different geodesic segments on a fixed time slice.
The power-law $\gamma = \gamma(\epsilon, \tilde \epsilon)$ defines the asymptotic behavior of the corresponding $n$-point correlation function while the conformal block starts with the constant term. As the  classical action is computed in bulk variables the exact correspondence in \eqref{block-action} assumes a  conformal transformation form the cylinder to the plane \cite{Hijano:2015rla}.

To clarify the proposed identification of the graphs and the respective quantities we explicitly focus on the $n=5$ case. Using the AGT correspondence \cite{Alday:2009aq} we compute the heavy-light classical conformal $5$-pt block. The corresponding worldline configuration in the bulk is described by a system of irrational equations which are too difficult to solve exactly. It  properly reflects the complexity of finding closed expressions for (classical) conformal blocks  \cite{Litvinov:2013sxa,Perlmutter:2015iya}. Instead, to solve the equation system we propose to use a series expansion method. Starting with a known exact (seed) solution to the equation system and expanding around the seed  with respect to some deformation parameter one finds the perturbative solution. As the seed solution we use a 5-pt block with one of the fields taken to be the unit operator what  precisely corresponds the 4-pt block considered previously in \cite{Fitzpatrick:2014vua,Hijano:2015rla}.

The next sections discuss  bulk/boundary realizations technically.
The exposition is organized as follows. In Sec. \bref{sec:boundcomp} we consider the AGT representation of $n$-point conformal blocks. In particular, in Sec. \bref{sec:fivepoint}
we explicitly compute the classical $5$-point conformal block. Then, in Sec. \bref{sec:worldline} we switch to the bulk analysis and discuss general properties of a probe particle worldlines in the background geometry. In this paper we consider the case of conical deficit only ($\alpha^2>0$). In Sec. \bref{sec:5line} we explicitly formulate the system of equations underlying the respective five-line graph in the bulk. Sec. \bref{sec:deformation} discusses the perturbation method that  treats the $5$-pt case as a deformation of the $4$-pt case. In Sec. \bref{sec:pert} we find an exact formula for the corresponding worldline action and compare it with the boundary results. In the last section \bref{sec:multi} we propose a multi-line generalization of the approach supported in the 5-line case. Sec. \bref{sec:conclusion}
contains our conclusions and outlooks.

\section{Boundary computation}
\label{sec:boundcomp}
The boundary computation is reduced to the analysis of the classical conformal block in the Virasoro CFT.
In this context there exist many different methods, each one having its own advantages and disadvantages. For instance, the elliptic recursion method can be easily combined with the semiclassical limit \cite{Zamolodchikov:1985ie}. For  generic four-point classical conformal block on the sphere  an explicit representation in terms of the regularized action evaluated on certain solution of the Painlev\'{e} VI equation is available \cite{Litvinov:2013sxa}.
Another method for computing the classical conformal block directly is the monodromy
method \cite{Zamolodchikov1986,Harlow:2011ny}. However, these methods require quite complicated independent analysis for each particular number of insertions in the correlator which is not easily generalized.

We find it instructive to adopt here another, the  so-called AGT method (see \cite{Alday:2009aq} and references therein for Liouville theory and \cite{Alkalaev:2014sma,Bershtein:2014qma} for Minimal Models) which gives a  simple uniform representation of the conformal blocks for arbitrary number of fields and for any genus.
 
\subsection{Combinatorial representation}
\label{sec:combrepr}

The AGT correspondence \cite{Alday:2009aq}   establishes a connection between $2d$ conformal field theories and the special class of $4d$ gauge theories in the so-called Omega background (for more details see \cite{Nekrasov:2002qd} and references therein). In particular, the correspondence allows to express the
CFT conformal  block functions in terms of the Nekrasov instanton partition functions for the gauge theories with the special matter fields content encoded in the dual pant decomposition diagram. These instanton partition functions are known explicitly, therefore using AGT leads to the explicit
results for arbitrary $n$-point conformal block. We note also that AGT is applicable  on a surface of arbitrary genus.

First, let us describe the AGT construction for the general  $n$-point conformal block\footnote{Here we discuss the construction 
of the conformal blocks on the sphere with the dual diagram having no closed loops. For higher genera, the diagrams associated to the block functions contain a number of loops and the AGT formulas require  simple modifications.} associated to the correlation function $\langle\Phi_{1}(z_{1},\bar{z}_{1})\dots\Phi_{n}(z_{n},\bar{z}_{n})\rangle$. 
Using projective invariance we fix three points  $z_{1}=0$, $z_{n-1}=1$, $z_{n}=\infty$, and  replace
\begin{equation}
\label{zq}
z_{i+1}=q_{i}q_{i+1}\dots q_{n-3}\quad\text{for}\quad1\leq i\leq n-3\;.
\end{equation}
The conformal block is given by the following series expansion
\begin{equation}\label{conformal-block-explicit}
    \mathcal{F}(q|\Delta,\tilde{\Delta},c)=
    1+\sum_{k}q_{1}^{k_{1}}q_{2}^{k_{2}}\dots q_{n-3}^{k_{n-3}}\,\mathcal{F}_{k}(\Delta,\tilde{\Delta},c),
\end{equation}
where $\Delta=\{\Delta_{1},\dots,\Delta_{n}\}$ is the set of external dimensions, 
$\tilde{\Delta}=\{\tilde{\Delta}_{1},\dots,\tilde{\Delta}_{n-3}\}$ is the set of the  intermediate dimensions and 
$q=\{q_1,\dots, q_{n-3}\}$.
The sum in \eqref{conformal-block-explicit} goes over all sets of positive integers, $k = \{k_1,...,k_{n-3}\}$.

Using the standard Liouville parametrization,
\be
\label{Liouville}
\ba{c}
\dps
\Delta_{i}=\frac{Q^{2}}{4}-P_i^{2}\;,
\qquad
\tilde{\Delta}_{j}=\frac{Q^{2}}{4}-\tilde{P}_j^{2}\;,\qquad c=1+6Q^{2}\;,\qquad Q=b+\frac{1}{b}\;,
\ea
\ee
the AGT representation of the $n$-point conformal block is given  as \cite{Alday:2009aq,Alba:2010qc}
\begin{equation}
\label{FF}
   \mathcal{F}(q|\Delta,\tilde{\Delta},c)=
    \prod_{r=1}^{n-3}\prod_{s=r}^{n-3}(1-q_{r}\dots q_{s})^{2(P_{r+1}-\frac{Q}{2})(P_{s+2}+\frac{Q}{2})}\,\,
    \mathcal{Z}(q|\Delta,\tilde{\Delta},c),
\end{equation}
where
\begin{equation}
\label{ZZ}
    \mathcal{Z}(q|\Delta,\tilde{\Delta},c)=
    1+\sum_{k}q_{1}^{k_{1}}q_{2}^{k_{2}}\dots q_{n-3}^{k_{n-3}}\,\mathcal{Z}_{k}(\Delta,\tilde{\Delta},c),
\end{equation}
and
\begin{multline}
\label{Zvac}
   \mathcal{Z}_{k}(\Delta,\tilde{\Delta},c)=
\sum_{\vec{\lambda}_{1},\dots,\vec{\lambda}_{n-3}}\!\!\!\!\!\frac{Z(P_{2}|P_1,\varnothing;\tilde{P}_{1},\vec{\lambda}_{1})Z(P_{3}|\tilde{P}_{1},\vec{\lambda}_{1};\tilde{P}_{2},\vec{\lambda}_{2})
   \cdots 
Z(P_{n-1}|\tilde{P}_{n-3},\vec{\lambda}_{n-3};P_n,\varnothing)}
{Z(\frac{Q}{2}|\tilde{P}_1,\vec{\lambda}_1;\tilde{P}_1,\vec{\lambda}_1) \cdots  
Z(\frac{Q}{2}|\tilde{P}_{n-3},\vec{\lambda}_{n-3};\tilde{P}_{n-3},\vec{\lambda}_{n-3})}.
\end{multline}
Here, the sum goes over $(n-3)$ pairs of Young tableaux $\vec\lambda_j =(\lambda_j^{(1)}, \lambda_j^{(2)})$ with the total number of cells $|\vec{\lambda}_{j}|\equiv |\lambda_j^{(1)}|+|\lambda_j^{(2)}|=k_{j}$.  The explicit form of  functions $Z$  reads
\be
\ba{c}
\dps
\label{ZZ}
    Z(P''|P',\vec{\mu};P,\vec{\lambda})=
\\
\\
\dps
\prod_{i,j=1}^{2}\,
    \prod_{s\in \lambda_{i}}\left(P''-E_{\lambda_{i},\mu_{j}}\bigl((-1)^j P'-(-1)^i P\bigl|s\bigr)+\frac Q2\right)
    \prod_{t\in \mu_{j}}\left(P''+E_{\mu_{j},\lambda_{i}}\bigl((-1)^i P-(-1)^j P'\bigl|t\bigr)-\frac Q2\right),  
\ea
\ee
where 
\begin{equation}
\label{E-def}
    E_{\lambda,\mu}\bigl(x\bigl|s\bigr)=x-b\,l_{\mu}(s)+b^{-1}(a_{\lambda}(s)+1).
\end{equation}
For a cell $s=(m,n)$ such that $m$ and $n$ label a respective row  and a column,
the arm-length function $a_{\lambda}(s) = (\lambda)_m-n$ and 
the leg-length function $l_{\lambda}(s) = (\lambda)^T_n - m$, where  
$(\lambda)_m$ is the length of  $m$-th row of the Young tableau $\lambda$,
and  $(\lambda)^T_n$ the height of the $n$-th column, where $T$ stands for a matrix transposition.

We note finally that the AGT equations  \eqref{FF}-\eqref{ZZ} give an efficient method for calculating
the coefficients of the series expansion \eqref{conformal-block-explicit} of the general conformal block function.

\subsection{Five-point classical conformal block} 
\label{sec:fivepoint}

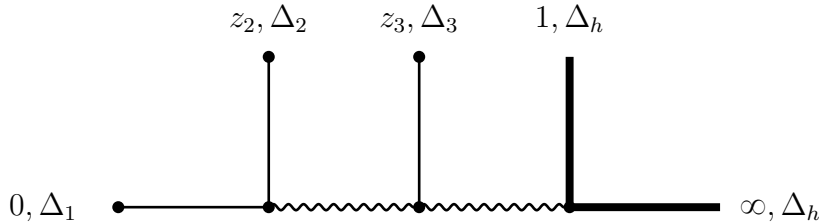
\begin{figure}[H]
\centering
\begin{tikzpicture}

\draw [line width=1pt] (30,0) -- (32,0);
\draw [line width=1pt] (32,0) -- (32,2);
\draw [smooth, tension=1.0, line width=1pt, decorate, decoration = {snake, segment length = 2mm, amplitude=0.4mm}] (32,0) -- (34,0);
\draw [line width=1pt] (34,0) -- (34,2);
\draw [smooth, tension=1.0, line width=1pt, decorate, decoration = {snake, segment length = 2mm, amplitude=0.4mm}] (34,0) -- (36,0);
\draw [line width=3pt] (36,0) -- (36,2);
\draw [line width=3pt](36,0) -- (38,0);


\draw (29,-0) node {$0, \Delta_1$};
\draw (32,2.5) node {$z_2, \Delta_{2}$};
\draw (34,2.5) node {$z_{3}, \Delta_{3}$};
\draw (36,2.5) node {$1, \Delta_{h}$};
\draw (38.8,0) node {$\infty, \Delta_h$};


\fill (30,0) circle (0.8mm);

\fill (32,0) circle (0.8mm);

\fill (34,0) circle (0.8mm);
\fill (32,2) circle (0.8mm);

\fill (36,0) circle (0.8mm);
\fill (34,2) circle (0.8mm);

\end{tikzpicture}
\caption{The $5$-point classical heavy-light conformal block. Two bold lines on the right represent heavy operators.}
\label{5block}
\end{figure}

Now we apply the above general result to $5$-point conformal block with the dual diagram depicted  in Fig. \bref{5block}.

\noindent Here, in terms of the parameters $q_1$ and $q_2$ \eqref{zq} the coordinates are 
\be
\label{zq5}
z_1 = 0 \;,
\qquad
z_2 = q_1 q_2\;,
\qquad
z_3 =q_2\;,
\qquad
z_4 = 1\;,
\qquad
z_5 = \infty \;.
\ee
Taking $n=5$ in the general representation \eqref{FF}-\eqref{ZZ} one finds 
\begin{equation}
   \mathcal{F}(q_1,q_2)=(1 - q_1)^{2 (P_2-\frac{Q}{2}) (P_3+\frac{Q}{2})} (1 - 
    q_1 q_2)^{2 (P_2-\frac{Q}{2}) (P_4+\frac{Q}{2})} (1 - 
    q_2)^{2 (P_3-\frac{Q}{2}) (P_4+\frac{Q}{2})}
   \mathcal{Z}(q_1,q_2),
\end{equation}
where 
\begin{equation}
\label{calZ}
   \mathcal{Z}(q_1,q_2)=
    1+\sum_{k_1,k_2}q_{1}^{k_{1}}q_{2}^{k_{2}}\,\mathcal{Z}_{k_1,k_2},
\end{equation}
and 
\begin{align}
\label{calZZ}
   \mathcal{Z}_{k_1,k_2}=\!\!\!\!
\sum_{\vec{\lambda}_{1},\vec{\lambda}_{2}}^{|\vec{\lambda}_{1,2}|=k_{1,2}}
\frac{Z(P_{2}|P_1,\varnothing;\tilde{P}_{1},\vec{\lambda}_{1})Z(P_{3}|\tilde{P}_{1},\vec{\lambda}_{1};\tilde{P}_{2},\vec{\lambda}_{2})
Z(P_{4}|\tilde{P}_{2},\vec{\lambda}_{2};P_5,\varnothing)}
{Z(\frac{Q}{2}|\tilde{P}_1,\vec{\lambda}_1;\tilde{P}_1,\vec{\lambda}_1)
Z(\frac{Q}{2}|\tilde{P}_{2},\vec{\lambda}_{2};\tilde{P}_{2},\vec{\lambda}_{2})}\;,
\end{align}
where on the lower levels the pairs of Young tableaux  $\vec{\lambda}=(\lambda^{(1)},\lambda^{(2)})$ with the total number of cells $l=|\vec{\lambda}|$ are
\be
\ba{l}
\dps
l=0:\quad\{(\varnothing,\varnothing)\}
\\
\\
\dps
l=1:\quad\{(\varnothing,\tableau{1}),(\tableau{1},\varnothing)\}
\\
\\
\dps
l=2:\quad\{(\varnothing,\tableau{2}),(\varnothing,\tableau{1 1}),(\tableau{1},\tableau{1}),(\tableau{2},\varnothing),(\tableau{1 1},\varnothing)\}
\\
\\
\dps
l=3:\quad\{(\varnothing,\tableau{3}),(\varnothing,\tableau{2 1}),(\varnothing,\tableau{1 1 1}),(\tableau{1},\tableau{2}),(\tableau{1},\tableau{1 1}),
\\
\\
\hspace{18mm}(\tableau{2},\tableau{1}),(\tableau{1 1},\tableau{1}),(\tableau{3},\varnothing), (\tableau{2 1},\varnothing), (\tableau{1 1 1},\varnothing)\} \;.
\\
\\
\ea
\ee

In what follows we are interested in the conformal block with dimensions 
\be
\label{PPPP}
P_4=P_5\;,
\qquad
P_1=P_2\;,
\qquad
\tilde{P}_1=\tilde{P}_2\;.
\ee
We find
\be
\label{calZt}
\mathcal{Z}(q_1,q_2|t)= 1+ \frac{(1 + b^2 - 2 b P_3) (1 + b^2 + 2 b P_3) (q_1 + q_2) }{8 b^2}t+ \cO(t^2)\;,
\ee
where $t$ is convenient formal parameter,  $t^m$ term takes into account contributions $q_1^{m_1}q_2^{m_2}$ with $m=m_1+m_2$. To reproduce the original function \eqref{calZ} one sets $t=1$ in \eqref{calZt} so that   $\mathcal{Z}(q_1,q_2)=\mathcal{Z}(q_1,q_2|1)$. Higher order expansion coefficients in $t$ needed for the subsequent analysis are not shown here, and can be directly read off the general formula \eqref{calZZ}.

\vspace{-3mm}

\paragraph{Classical 5-pt conformal block.}  Within the the Liouville parametrization  \eqref{Liouville} the limit $c\rightarrow 0$ can be equivalently understood as $b\rightarrow 0$, so that $b^2 = 6/c$. In particular, it is convenient to define classical dimensions as $\epsilon = 6\Delta/c$.  Then, the classical conformal block in \eqref{classblockdef} is given by 
\begin{equation}
\label{5classblock}
\mathcal{F}(q_1,q_2)= e^{-\frac{f(q_1,q_2)}{b^2}}\;,\qquad b\rightarrow 0\;.
\end{equation}
Equivalently, 
\begin{equation}
f(q_1,q_2)=-\lim_{b\rightarrow0} b^2 \ln \cF(q_1,q_2)\;.
\end{equation}
Liouville parameters $P_i$ are expressed via  conformal dimensions $\Delta_i$ by means of the following substitution 
\be
P_4=P_5=\sqrt{\frac{(b + 1/b)^2}{4} - \frac{\epsilon_h}{b^2}}\;,
\ee
\be
P_1=P_2=\sqrt{\frac{(b + 1/b)^2}{4} -\delta \frac{\epsilon_1}{b^2}}\;,
\ee
\be
\;\;\;\;\;\;\;\;P_3=\sqrt{\frac{(b + 1/b)^2}{4} - \delta \frac{\epsilon_3}{b^2}}\;,
\ee
\be
\tilde{P}_1=\tilde{P}_2=\sqrt{\frac{(b + 1/b)^2}{4} - \delta \frac{\tilde{\epsilon}_1}{b^2}}\;,
\ee
where $\delta$ is the lightness parameter \eqref{lightness}. Note that $P_{4,5}$ associated to heavy fields are of the zeroth order in $\delta$.

Following the  general discussion in the Introduction the expansion in $\delta$ \eqref{lightness} corresponds to the semiclassical expansion in the bulk with the first order contributions identified as in \eqref{block-action}. Hence, according \eqref{5classblock} we collect terms of order $b^{-2}$ in the expansion of  $\ln \cF(q_1,q_2)$ and then expand in $\delta$ up to first order.
Using the generating  functions in the formal variable $t$ one finds 
\be
f(q_1,q_2|t)=f_{\delta}(q_1,q_2|t) \delta +f_{\delta^2}(q_1,q_2|t) \delta^2+...\;,
\ee
where explicit form of the first two coefficients is
\be
\label{5ptdec}
\ba{l}
\dps
f_{\delta}(q_1,q_2|t)=-\frac{\epsilon_3 (q_1 + q_2)}{2}t +
\Big[\frac{(\epsilon_3-2 \tilde{\epsilon}_1) q_1 q_2}{4}- \frac{\epsilon_3(\epsilon_3+2) q_1^2}{16 \tilde{\epsilon}_1}   - 
\frac{\epsilon_3 q_2^2}{8} -\frac {2\epsilon_3 \epsilon_h q_2^2}{3} - \frac{\epsilon_3^2 q_2^2}{16 \tilde{\epsilon}_1} + \frac{\epsilon_3^2 \epsilon_h q_2^2}{
 4 \tilde{\epsilon}_1}\Big]t^2+
 \\
\\
\dps
+\Big[
-\frac{\epsilon_3 q_1^3}{24} - \frac{\epsilon_3^2 q_1^3}{16 \tilde{\epsilon}_1} - \frac{\epsilon_3 q_1^2 q_2}{8} + \frac{\epsilon_3^2 q_1^2 q_2}{16 \tilde{\epsilon}_1}
 - \frac{\epsilon_3 q_1 q_2^2}{8} + \frac{2 \epsilon_3 \epsilon_h q_1 q_2^2}{3} + \frac{\epsilon_3^2 q_1 q_2^2}{16 \tilde{\epsilon}_1} - \frac{\epsilon_3^2 \epsilon_h q_1 q_2^2}{4 \tilde{\epsilon}_1}   -  \frac{\epsilon_3 q_2^3}{24} - \frac{2 \epsilon_3 \epsilon_h q_2^3}{3} -
\\
\\
\dps 
 - \frac{\epsilon_3^2 q_2^3}{16 \tilde{\epsilon}_1} + \frac{\epsilon_3^2 \epsilon_h q_2^3}{4 \tilde{\epsilon}_1}\Big]t^3
+\cO(t^4)\;,
\ea
\ee
and
\be
\ba{l}
\dps
f_{\delta^2}(q_1,q_2|t)=
\dps
\Big[\frac{\epsilon_3^2 q_1^2}{16 } - \frac{2 \epsilon_3 \epsilon_1 q_1^2}{3} + \frac{\epsilon_3^2 \epsilon_1 q_1^2}
{ 4 \tilde{\epsilon}_1} - \frac{\epsilon_3 \tilde{\epsilon}_1 q_1^2}{6} + \frac{\epsilon_3^2 q_2^2}{16 } - \frac{
   \epsilon_3^2 \epsilon_h q_2^2}{3 } - \frac{\epsilon_3 \tilde{\epsilon}_1 q_2^2}{6 } + \frac{8 \epsilon_3 \epsilon_h \tilde{\epsilon}_1 q_2^2}{
   9 }\Big] t^2 +\cO(t^3)\;.
\ea
\ee
With the change 
\be
\epsilon_h=\frac{1-\alpha^2}{4}\;,
\ee
where parameter $\alpha^2 >0$ will be associated with the background metric generated by the heavy fields, in the first order in $\delta$ we find the expansion
\be
\label{5ptdec1}
\ba{l}
\dps
f_{\delta}(q_1,q_2|t)=\Big[-\frac{\epsilon_3 q_1}{2} - \frac{\epsilon_3 q_2}{2}\Big] t 
+\Big[-\frac{\epsilon_3 q_1^2}{8} - \frac{\epsilon_3^2 q_1^2}{16 \tilde{\epsilon}_1} + \frac{\epsilon_3 q_1 q_2}{4} - \frac{\tilde{\epsilon}_1 q_1 q_2}{2}
 - \frac{7 \epsilon_3 q_2^2}{24} + \frac{\epsilon_3 q_2^2 \alpha^2}{6} - \frac{\epsilon_3^2 q_2^2 \alpha^2}{16 \tilde{\epsilon}_1}\Big] t^2
\\
\\
\dps 
  +\Big[-\frac{\epsilon_3 q_1^3}{24} - \frac{\epsilon_3^2 q_1^3}{16 \tilde{\epsilon}_1} - \frac{\epsilon_3 q_1^2 q_2}{8} + \frac{\epsilon_3^2 q_1^2 q_2}{16 \tilde{\epsilon}_1} + \frac{\epsilon_3 q_1 q_2^2}{24} - \frac{5 \epsilon_3 q_2^3}{24} - 
   \frac{\epsilon_3 q_1 q_2^2 \alpha^2}{6} + \frac{\epsilon_3^2 q_1 q_2^2 \alpha^2}{16 \tilde{\epsilon}_1} + 
   \frac{\epsilon_3 q_2^3 \alpha^2}{6} - \frac{\epsilon_3^2 q_2^3 \alpha^2}{16 \tilde{\epsilon}_1}\Big] t^3
\\
\\
\dps 
+\cO(t^4)\;.
\ea
\ee

The bulk computation procedure discussed in the next sections allows to reconstruct the classical conformal block  using a series expansion  around some exact seed solution to the bulk equations.  In Sec. \bref{sec:deformation} we will explain this procedure taking as a seed solution the  five-point classical block with the unity operator insertion $\Phi_3  = \mathbb{I}$. This means that the deformation parameter is identified with the conformal dimension $\epsilon_3$ and in the limit $\epsilon_3 = 0$  the 5-pt block goes to the $4$-pt one. Hence, to compare with the results of the bulk computation  we will need an expansion of the 5-pt classical conformal block  with respect to the parameter $\epsilon_3$, \textit{i.e.},
\be
\label{pertu}
f_{\delta}(q_1,q_2|t)=f^{(0)}_{\delta}(q_1,q_2|t)+\epsilon_3 f^{(1)}_{\delta}(q_1,q_2|t)+\epsilon_3^2 f^{(2)}_{\delta}(q_1,q_2|t)+...\;.
\ee
Here, the leading term $f^{(0)}_{\delta}(q_1,q_2|1)$ is identified with the 4-pt classical conformal block, while the sub-leading terms perturbatively reconstruct the 5-pt classical conformal block. The explicit form of the first two terms read off from \eqref{5ptdec1} is
\be
\ba{l}
\dps
f^{(0)}_{\delta}(q_1,q_2|t)=
-\frac{1}{2} \tilde{\epsilon}_1 q_1 q_2 t^2 + 
 \frac{1}{48} \big(-4 \epsilon_1 q_1^2 q_2^2 - 10 \tilde{\epsilon}_1 q_1^2 q_2^2 + 4 \epsilon_1 q_1^2 q_2^2 \alpha^2 +
     \tilde{\epsilon}_1 q_1^2 q_2^2 \alpha^2\big) t^4 
\\
\\
\dps      
\hspace{3.2cm}  +   \frac{1}{48} \big(-4 \epsilon_1 q_1^3 q_2^3 - 6 \tilde{\epsilon}_1 q_1^3 q_2^3 + 4 \epsilon_1 q_1^3 q_2^3 \alpha^2 + 
   \tilde{\epsilon}_1 q_1^3 q_2^3 \alpha^2\big) t^6 +\cO(t^8)\;,
\ea
\ee
and
\be
\ba{l}
\dps
f^{(1)}_{\delta}(q_1,q_2|t)=-\frac{1}{2}(q_1+ q_2) t + 
 \frac{1}{24} \big(-3 q_1^2 + 6 q_1 q_2 - 7 q_2^2 + 4 q_2^2 \alpha^2\big) t^2 
\\
\\
\dps  
\hspace{3.2cm}+\frac{1}{24} \big(-q_1^3 - 3 q_1^2 q_2 + q_1 q_2^2 - 5 q_2^3 - 4 q_1 q_2^2 \alpha^2 + 
    4 q_2^3 \alpha^2\big) t^3+\cO(t^4)\;.
\ea
\ee
As expected, function  $f^{(0)}_{\delta}(q_1,q_2|t)$ depends on the combination $q_1q_2$ only which is identified with the coordinate $z_2$, cf. \eqref{zq5}.
Setting $t=1$ one can recognize in the above  series expression the following functions 
\be
\label{finblock1}
\ba{l}
\dps
f^{(0)}_{\delta}(q_1,q_2)=
2 \epsilon_1 \ln\Big[-\frac{2 \sinh[\frac{\alpha \ln[1 - q_1 q_2 ]}{2}]}{\alpha q_1 q_2 } \Big]- 
 \tilde{\epsilon}_1 \ln\Big[-\frac{4  \tanh[\frac{\alpha \ln[1 - q_1 q_2]}{4}]}{ \alpha q_1 q_2  }\Big] + \epsilon_1 \ln[1 - q_1 q_2 ]\;,
\ea
\ee
and
\be
\label{finblock2}
\ba{l}
\dps
f^{(1)}_{\delta}(q_1,q_2)=
\ln\Big[\frac{\sinh[\frac{\alpha (\ln[1 - q_1 q_2 ] - 2 \ln[1 - q_2   ])}{2}]}{\alpha q_2 }\Big] + \ln[1 - q_2  ]\;.
\ea
\ee
After a particular conformal transformation of the coordinates these are the expansion coefficients that are seen on the bulk side in Sec. \bref{sec:pert}. 

A few comments are in order. Firstly, the choice of conformal dimensions \eqref{PPPP} is not assumed to be clear and requires some explanation. 
As we already discussed, the bulk computations corresponding to our
boundary configuration  (discussed in the next sections) rely on the specially developed 
perturbation procedure around known exactly seed solution. So, in order to find explicitly general 5-point conformal blocks we have to fix a number of perturbation parameters. In our case we consider one possible choice corresponding to the value $\epsilon_3=0$   \eqref{pertu}.
On the general physical grounds (namely, taking into account fusion rules) in this case we forced to fix $\tilde{\epsilon}_1=\tilde{\epsilon}_2$. This explains our choice \eqref{PPPP}.
More general conformal block within the framework of this perturbation procedure  can be reconstructed order by order 
with respect to each perturbation parameter. Thus, to evaluate the conformal block with
$\tilde{\epsilon}_1\neq\tilde{\epsilon}_2$ we have to introduce additional small parameter $\delta'=\tilde{\epsilon}_1-\tilde{\epsilon}_2$, and  develop corresponding perturbation
theory which is based on the same idea, and so represents not conceptual but technical difference.

Secondly, we note that the AGT method used here allows to get only series expansion of the conformal block functions. So that, essentially what we perform is the check
that the series expansion of the resulting expression obtained in the bulk computation gives exactly the coefficients of the classical conformal block
calculated up to rather high order $\cO(q_1^mq_2^n)$. Thus,  the bulk computation of Sec. \bref{sec:pert} allows to derive the exact result, while the boundary computation allows only to  conjecture this expression and check the lower level coefficients comparing with the bulk. It would be interesting to actually derive this exact result from the boundary point of view.

\section{Worldline approach}
\label{sec:worldline}

Let us consider a massive  point particle moving in the 
background  with a conical defect. In the cylindrical coordinate system $x^\mu = (t, \phi, \rho)$ the metric  $g_{\mu\nu}(x)$ can  be read off from the interval    
\be
\label{metric}
ds^2  = \frac{\alpha^2}{\cos^2 \rho}\Big( - dt^2 +\sin^2\rho d\phi^2 +\frac{1}{\alpha^2} d\rho^2 \Big)\;,
\ee
where  $\alpha^2 >0$ parameterizes an angle deficit. The physical singularity is placed at $\rho = 0$, where the Riemann tensor component $R_{\rho\phi\rho\phi}$ blows up. The conformal boundary corresponds to points $\rho = \pi/2$. To approach the boundary we use the regularization $\cos \rho = \Lambda^{-1}$ at $\Lambda \rightarrow \infty$. 

Particle propagation between initial and final positions  is described by  the worldline action   
\be
\label{OPA}
S = m \dps\int_{\lambda^{'}}^{\lambda^{''}} d\lambda \,\sqrt{g_{\mu\nu}(x) \dot{x}^\mu\dot{x}^\nu}\equiv m\int_{\lambda^{'}}^{\lambda^{''}} d\lambda \, \sqrt{g_{tt} \dot{t}^2+g_{\phi\phi} \dot{\phi}^2+g_{\rho\rho} \dot{\rho}^2}\;, 
\ee 
where $m$ is the mass of a particle, $\lambda$ is the  evolution parameter, and $\dot{x}^\mu = d x^{\mu}/d \lambda$. In general, ending points are parameters of the theory and the variation of the on-shell action reads
\be
\label{variation_bound}
\delta S  = p_\mu^{''}\delta x^{'' \mu} - p_\mu^{'}\delta x^{' \mu}\;, 
\ee
where $p^{'}_\mu$ and $p^{''}_\mu$ are momenta in the initial and final positions, $\delta x^{' \mu}$ and $\delta x^{'' \mu}$ are respective coordinate variations.   
The corresponding Euler-Lagrange equations of motion are the geodesic equation provided that  $\lambda$ is identified with a length of the path. In this case one arrives at  the normalization condition
\be 
\label{properpar}
 |\dot{x}|  \equiv \sqrt{g_{\mu\nu}(x) \dot{x}^\mu\dot{x}^\nu} = 1\;.
\ee

The bulk dynamics can be reduced to a constant time disk with polar coordinates $\rho$ and $\phi$. Indeed,  $t$ and $\phi$ are cyclic coordinates resulting in the conservation laws,
$\dot{p}_t = 0$ and $\dot{p}_\phi = 0$,    
where $p_t = g_{tt}\dot{t}$ and $p_\phi = g_{\phi\phi} \dot{\phi}$ are the corresponding  momenta. Choosing   a particular value $p_t = 0$ one arrives at $\dot{t} = 0$  $\rightarrow$ $t(\lambda) = const$. From now on we choose a constant time slice  $t=0$. Taking into account the normalization condition \eqref{properpar} integration constants can be identified with initial and final radial positions and the value of the conserved angular momentum $p_\phi$. Solving equations of motion explicitly is superfluous as we need just an action value  evaluated on a given path. Using \eqref{properpar} one finds that the action is a length of the path 
\be
\label{actlambda}
S = \int_{\lambda^{'}}^{\lambda^{''}} d \lambda = \lambda^{''} - \lambda^{'}\;. 
\ee

A useful trick is that the normalization condition \eqref{properpar} is sufficient to express a proper parameter $\lambda$ as a function of radius and  angular momentum values \cite{Hijano:2015rla}. For $\dot{t} = 0$ the condition \eqref{properpar}  can be cast into the form 
\be
\label{geod}
\frac{1}{\cos^2\rho} \dot{\rho}^2 + \frac{p_{\phi}^2}{\alpha^2}\cot^2\rho = 1\;. 
\ee 
from which it follows that the radial velocity is expressed as 
\be
\label{rhorad}
\dot \rho = \pm \cos\rho\, \sqrt{1 - \frac{p_\phi^2}{\alpha^2}\cot^2 \rho}\;\;,
\ee
where the overall sign depends on the direction of the $\lambda$ flow. The minimal radial distance between the particle  path and the singularity is therefore
given by  $\tan^2 \rho_{min} = \Big(\frac{p_\phi}{\alpha}\Big)^2$. 
Obviously, the maximal radial distance corresponds to the point $\rho_{max} = \pi/2$ located on the boundary.  Changing variables as $y = \cot^2 \rho$ at $\dot \rho \geq 0$, and introducing notation 
\be
\label{s}
s = \frac{|p_\phi|}{\alpha}\;,
\ee
equation \eqref{geod} can be directly integrated to yield the on-shell action 
\be
\label{lambda}
S = \ln \frac{ \sqrt{\eta}}{\sqrt{1+\eta} +  \sqrt{1 - s^2 \eta}}\,\Bigg|_{\eta^{'}}^{\eta^{''}}\;,
\ee
where $\eta^{'} = \cot^2 \rho^{'}$ and $\eta^{''} = \cot^2 \rho^{''}$ are initial/final radial positions. Parameter $s$  is an integration constant that defines  a particular form of the geodesic segment.   

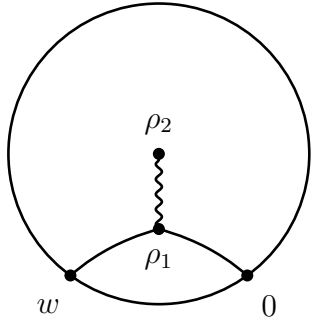
\begin{figure}[H]
\centering
\begin{tikzpicture}[line width=1pt]
\draw (0,0) circle (2.0cm);

\foreach \a in {1,2,...,40}{
\draw (\a*360/40: 2.0cm) coordinate(K\a){};
\draw (\a*360/40: 1.4cm) coordinate(L\a){};
\draw (\a*360/40: 2.5cm) coordinate(M\a){};
\draw (\a*360/40: 1cm) coordinate(I\a){};}


\draw plot [smooth, tension=1.0, line width=1pt] coordinates {(K34) (L33) (I30)};
\draw plot [smooth, tension=1.0, line width=1pt] coordinates {(K26) (L27) (I30)};



\draw [smooth, tension=1.0, line width=1pt, decorate, decoration = {snake, segment length = 2mm, amplitude=0.4mm}] (I30)  -- (0,0);

\fill (I30) circle (0.8mm);
\fill (K34) circle (0.8mm);
\fill (K26) circle (0.8mm);
\fill (0,0) circle (0.8mm);

\draw (M34) node {$0$};
\draw (M26) node {$w$};
\draw (0,0.4) node {$\rho_2$};
\draw (0,-1.4) node {$\rho_1$};

\end{tikzpicture}
\caption{Radial and arc segments. The graph corresponds to the classical conformal block with two heavy fields, two light fields of equal dimensions (the arc), and one extremely light intermediate field (the radial line) \cite{Hijano:2015rla}.  } 
\label{line}
\end{figure}

The simplest case of a geodesic segment is the radial line starting (or ending, depending on the $\lambda$ flow direction) at the singularity point $\rho_2 = 0$, see Fig. \bref{line}. In this case,  the angular momentum  $p_\phi$ vanishes so that  $s=0$.  After some simple algebra,  one finds from \eqref{lambda} the radial length  
$S_{rad} =  -\ln \tan (\frac{\rho_1}{2}+\frac{\pi}{4})$. We see that $S_{rad}$ is finite implying that a particle reaches the singularity within a finite time period. One interprets the falling into the singularity as a cubic vertex of the two heavy operators and a light operator represented by a probe. For the further purpose we find a length of the radial line for $\rho_1 = \arccos \sin (\alpha w/2)$:
\be
\label{rad}
S_{rad} =  - \ln \tan \frac{\alpha w}{4}\;. 
\ee

For the geodesic arc connecting two boundary points $\phi = 0$ and $\phi = w$  t11he angular momentum $p_\phi$ is not vanishing $s = \cot \frac{\alpha w}{2}$. From \eqref{lambda} it follows \cite{Roberts:2012aq,Asplund:2014coa,Hijano:2015rla} that the length of the arc is given by
\be
\label{arc}
S_{arc} = \ln \Big[\sin\frac{\alpha w}{2}\Big] + \ln 2\Lambda\;.
\ee
In particular, $S_{arc}$ diverges at $\Lambda \rightarrow \infty$ so that it takes an infinite time to reach the boundary. Note that $\rho_1 = \arccos \sin \frac{\alpha w}{2}$ chosen to compute \eqref{rad} corresponds to the zero value of the radial velocity, or the minimal distance according to formula \eqref{rhorad}. 
From the graph in Fig. \bref{line} it is clear that the minimal distance is given by \eqref{rad}.

\section{Five-particle configuration}
\label{sec:5line}

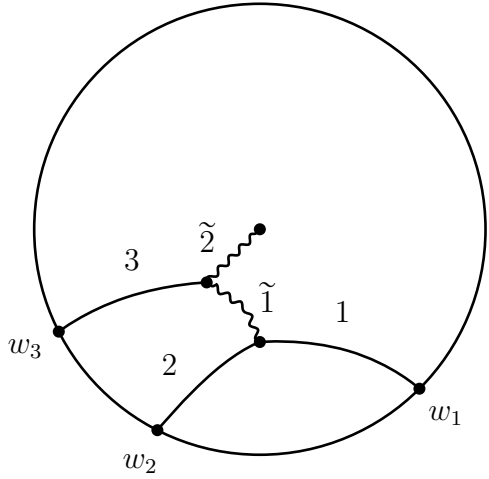
\begin{figure}[H]
\centering
\begin{tikzpicture}[line width=1pt]
\draw (0,0) circle (3cm);

\foreach \a in {1,2,...,40}{
\draw (\a*360/40: 3.5cm) coordinate(N\a){};
\draw (\a*360/40: 3cm) coordinate(K\a){};
\draw (\a*360/40: 2cm) coordinate(L\a){};
\draw (\a*360/40: 1.5cm) coordinate(I\a){};
\draw (\a*360/40: 1.1cm) coordinate(J\a){};
\draw (\a*360/40: 1cm) coordinate(M\a){};
;}


\draw plot [smooth, tension=1.0, line width=1pt] coordinates {(K35) (L34) (I30)};
\draw plot [smooth, tension=1.0, line width=1pt] coordinates {(K27) (L28) (I30)};
\draw plot [smooth, tension=1.0, line width=1pt] coordinates {(K23) (L23) (M25)};


\draw [smooth, tension=1.0, line width=1pt, decorate, decoration = {snake, segment length = 2mm, amplitude=0.4mm}] (M25)  -- (0,0);


\draw  [smooth, tension=1.0, line width=1pt, decorate, decoration = {snake, segment length = 2mm, amplitude=0.4mm}]  (I30) -- (J29) -- (M25);

\fill (I30) circle (0.8mm);
\fill (M25) circle (0.8mm);

\fill (K35) circle (0.8mm);
\fill (K27) circle (0.8mm);
\fill (K23) circle (0.8mm);
\fill (0,0) circle (0.8mm);

\draw (1.1,-1.1) node {$1$};
\draw (-1.2,-1.8) node {$2$};
\draw (-1.7,-0.4) node {$3$};
\draw (-0.7,-0.1) node {$\b$};
\draw (0.1,-0.9) node {$\a$};
\draw (N23) node {$w_3$};
\draw (N27) node {$w_2$};
\draw (N35) node {$w_1$};

\end{tikzpicture}
\caption{Five-particle  graph. Solid lines $1,2,3$ represent external particles, wavy lines $\a,\b$ represent intermediate particles. The angles are measured clockwise. In practice, we set $w_1=0$.} 
\label{5bulk}
\end{figure}

Consider now the five-line graph on Fig. \bref{5bulk} which is the $n=5$ case of the general graph in Fig. \bref{bulk}. 

\noindent The corresponding particle action reads 
\be
\label{act5}
S = \sum_I \epsilon_I S_I\;,
\qquad
I = 1,2,3,\a,\b\;,
\ee
where each component is given by \eqref{OPA}. Initial/final positions $\lambda^{'}$ and $\lambda^{''}$ correspond to various nodes in Fig. \bref{5bulk} including the singularity point, two vertices, three boundary attachments. It is supposed that the singularity point and  boundary attachments are fixed parameters of the theory. There is no loss of generality in supposing that $w_1 = 0$. From the boundary perspective it is achieved doing a conformal map that moves a position of the first external operator $z_1 \rightarrow 1$.
Positions of the vertices are  floating according to the minimal action principle.

From the normalization condition \eqref{properpar} it follows  $S_I = S_I^{''} - S_I^{'}$, and final/initial lengths are functions of the boundary points $w_2,w_3$, classical dimensions $\epsilon_1, \epsilon_2, \epsilon_3$ and $\tilde \epsilon_1, \tilde \epsilon_2$, and the metric parameter $\alpha$ , \textit{i.e.},
$S_I^{'} = S_I^{'}(w|\alpha, \epsilon)$ and  $S_I^{''} = S_I^{''}(w|\alpha, \epsilon)$. 
The total action \eqref{act5} is then $S = S(w|\alpha, \epsilon)$.

Let us consider each of  two vertices. In these points proper parameter $\lambda$ can be chosen to be increasing  away from the vertex.  Then, denoting the vertex coordinates as $ x_{1}^\mu$ and $ x_2^\mu$ along with the corresponding deviation $\delta  x_1^\mu$ and $\delta   x_2^\mu$ which are the same for all incoming lines, and using variation formula  \eqref{variation_bound} one arrives at two equilibrium  conditions $P_{1}{}_\mu \delta x_1^\mu =0$ and $ P_2{}_\mu \delta  x_2^\mu = 0$, where $P_\mu$ is a total momentum of lines incoming a given vertex. Explicitly, there are two following conditions.

\begin{itemize}

\item First vertex $\a -1 - 2$. The equilibrium condition reads 
\be
\label{a12}
\big(\tilde\epsilon_1 \tilde p_\mu^1 + \epsilon_1 p_\mu^1+\epsilon_2 p_\mu^2\big)\;\Big|_{x=x_1}  = 0\;,
\ee
where $x_1$ stands for  coordinates  of the first vertex.

\item Second vertex $\a -\b -  3$. The equilibrium condition reads
\be
\label{3ba}
\big(\tilde\epsilon_1 \tilde p_\mu^1+\tilde \epsilon_2 \tilde p_\mu^2 +\epsilon_3 p_\mu^3 \big)\;\Big|_{x=x_2}  = 0\;,
\ee
where $x_2$ stands for  coordinates of the second vertex.

\end{itemize}

\noindent Note that the action variations at the boundary attachments $w_2$ and $w_3$ are not vanishing. 
Instead, these are 
\be
\ba{c}
\label{wz}
\delta_{w_2}S(w_2,w_3|\alpha, \epsilon) = p_\phi^2(w_2,w_3|\alpha, \epsilon)\delta w_2\;,
\\
\\
\delta_{w_3} S(w_2,w_3|\alpha, \epsilon) = p_\phi^3(w_2,w_3|\alpha, \epsilon)\delta w_3\;,
\ea
\ee
where functions $p_\phi^2$ and $p_\phi^3$ are the angular momenta of external lines 2 and 3 attached to the corresponding points. Knowing these momenta explicitly one can integrate the equation system \eqref{wz} to find  the action $S =  S(w_2,w_3|\alpha, \epsilon)$ explicitly. In   the 4-pt case the system is given by a single ordinary differential equation and can be integrated explicitly  \cite{Hijano:2015rla}. However, for many-particle configurations a number of boundary attachments increases thereby making the system  \eqref{wz}  a partial differential equation system. Note that equations \eqref{wz} define the so-called accessory parameters for the classical conformal blocks which are identified here with the angular momenta \cite{Litvinov:2013sxa,Hijano:2015rla}.     

The other way around is to compute the  action  explicitly recalling that it is defined as a weighted sum of the path lengths \eqref{act5}. To this end, using the equilibrium conditions one explicitly finds positions of the vertices and expresses all angular momenta as functions of boundary attachments points $w_2$ and $w_3$. Summing up all the lengths which are now functions of $w_{2,3}$  one eventually arrives at  the sought-for action function $S = S(w_2,w_3|\alpha, \epsilon)$.  

\subsection{Vertex analysis}

Below we study the component form of the equilibrium conditions \eqref{3ba} and \eqref{a12}. Since the dynamics is reduced to a fixed time disk both conditions trivialize for $\mu = t$. The radial line $\b$ has vanishing angular momentum, $\tilde p_\phi^{2} =0$.

\vspace{-3mm}

\paragraph{Equilibrium  equations.} For the vertex $\a-\b-3$, the   components $\mu = \phi, \rho$ of the equilibrium condition \eqref{3ba} are given by 
\be
\tilde \epsilon_1 \tilde p_\phi^1+\epsilon_3 p_\phi^3 = 0\;,
\qquad
\epsilon_3 p_\rho^3 + \tilde \epsilon_1 \tilde p_\rho^1+\tilde \epsilon_2 \tilde p_\rho^2  = 0\;,
\ee
or, using the definition $p_\rho = g_{\rho\rho}\dot \rho$ and formula \eqref{rhorad}, one finds
\be
\label{lin1}
\epsilon_3 p_\phi^3 + \tilde \epsilon_1 \tilde p_\phi^1 = 0\;,
\ee
\be
\label{vertcoord1}
\epsilon_3\,\sqrt{1 - \frac{(p_\phi^3)^2}{\alpha^2 }\cot^2 \rho_2}  + 
\tilde \epsilon_1\,\sqrt{1 - \frac{(\tilde p_\phi^1)^2}{\alpha^2 }\cot^2 \rho_2} - \tilde \epsilon_2 = 0\;. 
\ee
Here $\rho_2$ is the radial coordinate of the $\a-\b-3$ vertex, and the radial momentum $\dot{\rho}_{(b)}<0$ since the proper  parameter $\lambda$ increases away from the vertex.  

For the vertex $\a - 1 - 2$, the  components $\mu = \phi, \rho$ of the equilibrium condition 
\eqref{a12} are given by 
\be
\label{lin2}
\tilde \epsilon_1 \tilde p_\phi^1 + \epsilon_1 p_\phi^1 + \epsilon_2 p_\phi^2= 0\;,
\ee
\be
\label{vertcoord2}
- \tilde \epsilon_1\,\sqrt{1 - \frac{(\tilde p_\phi^1)^2}{\alpha^2 }\cot^2 \rho_1}  + 
\epsilon_1\,\sqrt{1 - \frac{(p_\phi^1)^2}{\alpha^2 }\cot^2 \rho_1} 
+ \epsilon_2\,\sqrt{1 - \frac{(p_\phi^2)^2}{\alpha^2 }\cot^2 \rho_1} = 0\;. 
\ee
Here $\rho_1$ is the radial coordinate of the $\a - 1 - 2$ vertex and   $\dot{\rho}_{(\a)} < 0$, $\dot{\rho}_{(2)}>0$, and $\dot{\rho}_{(1)} > 0$ since the proper  parameter $\lambda$ increases away from the  vertex.  

\vspace{-3mm}

\paragraph{Independent integration constants.} The conserved angular momenta play the role of integration constants. In our case, there are  five 
angular momenta subjected to three independent constraints. Namely, as the radial line $\b$ has vanishing angular velocity and taking \eqref{lin1} and \eqref{lin2} into account one finds
\be
\tilde p_\phi^2 = 0\;, 
\qquad
\epsilon_3 p_\phi^3 + \tilde\epsilon_1 p_\phi^a = 0\;,
\qquad
\tilde \epsilon_1 \tilde p_\phi^1 + \epsilon_1 p_\phi^1 + \epsilon_2 p_\phi^2= 0\;.
\ee 
Using \eqref{s}  the above relations can be represented as follows
\be
\label{rels}
\tilde s_2 = 0\;,
\qquad
\epsilon_3 s_3 - \tilde \epsilon_1 \tilde s_1 = 0\;,
\qquad
\epsilon_1 s_1 - \epsilon_2 s_2 - \tilde \epsilon_1 \tilde s_1 = 0\;.
\ee
Note that quantities $s_I$ are non-negative and therefore the relative signs are fixed according to  the slopes of worldlines on  Fig. \bref{5bulk}. 

On the other hand, equations \eqref{vertcoord1} and \eqref{vertcoord2} can be used to find $\tan\rho_1$ and $\tan \rho_2$ as functions of two independent angular momenta (integration constants).

\vspace{-3mm}

\paragraph{Vertex $\a-\b - 3$ radial position.} The consideration here is similar to that one for  the $4$-pt block. From \eqref{vertcoord1} we have
\be
\label{vertcoord11}
\epsilon_3\sqrt{1-s_3^2 \eta_2} + \tilde \epsilon_1\sqrt{1-\tilde s_1^2 \eta_2} = \tilde \epsilon_2\;,
\ee
where angular parameters  $s_{3}$ and $\tilde s_1$ are defined according to \eqref{s}, and
\be
\label{etabar}
 \eta_2 = \cot^2{\rho_2}\;.
\ee
Solving equation  \eqref{vertcoord11} for $\eta_2$ one finds
\be
\label{1tan}
\eta_2  = -
\frac{(\epsilon_3 - \tilde \epsilon_1 - \tilde \epsilon_2) (\epsilon_3 + \tilde \epsilon_1 - \tilde \epsilon_1) 
(\epsilon_3 - \tilde\epsilon_1 + \tilde\epsilon_2) (\epsilon_3 + \tilde\epsilon_1 + \tilde\epsilon_2)}{4\, \epsilon_3^2\,\tilde\epsilon_2^2\, s_3^2}\equiv  \frac{\tau^2}{s_3^2}\;,
\ee
so that $\tau^2$ is a function of the conformal dimensions only 
\be
\label{tau}
\tau^2  = -
\frac{(\epsilon_3 - \tilde\epsilon_1 - \tilde\epsilon_2) (\epsilon_3 + \tilde\epsilon_1 - \tilde\epsilon_2) 
(\epsilon_3 - \tilde\epsilon_1 + \tilde\epsilon_2) (\epsilon_3 + \tilde\epsilon_1 + \tilde\epsilon_2)}{4\, \epsilon_3^2\,\tilde\epsilon_2^2}\;.
\ee
Function $\tau^2$ can be represented as $\tau^2  = \mu^2/ \epsilon_3^2$, where 
$\mu^2$ already appeared in the 4-pt case \cite{Hijano:2015rla},
\be
\label{mu}
\mu^2 = \frac{\epsilon_{3}^2 + \tilde\epsilon_1^2 - \tilde\epsilon_2^2/2}{2} - \frac{(\epsilon_3^2 - \tilde\epsilon_1^2)^2}{4\tilde\epsilon_2^2}\;.
\ee

\vspace{-3mm}

\paragraph{Vertex $\a - 1-2$ radial position.} Equation \eqref{vertcoord2} can be cast into the form 
\be
\label{vertcoord21}
\epsilon_1\sqrt{1-s_1^2 \eta_1} + \epsilon_2\sqrt{1-s_2^2 \eta_1} = \tilde\epsilon_1\sqrt{1-\tilde s_1^2  \eta_1}\;,
\ee
where momenta $s_{1,2}$ and $\tilde s_1$ are defined according to \eqref{s}, and $\eta_1$ is given by 
\be
\label{etatilde}
 \eta_1 = \cot^2{\rho_1}\;.
\ee
The general solution to \eqref{vertcoord21} reads
\be
\label{solution}
 \eta_1  =  \frac{1 - \sigma^2 }{s_1^2+s_2^2 -2 s_1s_2 \sigma  } \;,
\qquad\qquad 
\sigma = \frac{\epsilon_1^2 +\epsilon_2^2 - \tilde\epsilon_1^2}{2\epsilon_1\epsilon_2}\;.
\ee
Equivalently,  
\be
\label{cottilde}
 \eta_1  =\frac{(\epsilon_1 -\epsilon_2 - \tilde\epsilon_1) (\epsilon_1 + \epsilon_2 - \tilde\epsilon_1) 
(\epsilon_1 - \epsilon_2 + \tilde\epsilon_1) (\epsilon_1 + \epsilon_2 + \tilde\epsilon_1)}{4\epsilon_1 \epsilon_2\big( s_1 s_2(\epsilon_1^2 +\epsilon_2^2 - \tilde\epsilon_1^2) -\epsilon_1 \epsilon_2 (s_1^2 +s_2^2)\big)}\;,
\ee
cf. \eqref{1tan}.

It is worth noting that parameter functions  $\tau = \tau(\epsilon, \tilde\epsilon)$ \eqref{tau} and $\sigma = \sigma(\epsilon, \tilde\epsilon)$ \eqref{solution} are homogeneous functions of the conformal dimensions.
We note also  that both $\eta_1$ and $\eta_2$ contain  a classical "fusion polynomial" factor 
\be
\label{fusion}
\Pi(\epsilon_I, \epsilon_J, \epsilon_K) = (\epsilon_I - \epsilon_J - \epsilon_K)(\epsilon_I + \epsilon_J - \epsilon_K)(\epsilon_I - \epsilon_J + \epsilon_K)(\epsilon_I + \epsilon_J + \epsilon_K)\;,
\ee
where $I,J,K = 1,2,3,\a,\b$. There are two useful propositions. 

\begin{itemize}

\item From   $\Pi(\epsilon_I, \epsilon_J, \epsilon_K) \leq  0 $ it follows that
\be
\label{lem1}
\epsilon_I \leq  \epsilon_J + \epsilon_K\;, \
\qquad I \neq J \neq K\;.
\ee 
The proof is straightforward. In particular, it implies that $\eta_2 \geq 0$ and $\eta_1 \geq 0$. The first inequality is obvious from the definition \eqref{1tan}, while to show the second one we recall that $\epsilon_1 s_1 \geq \epsilon_2 s_2$ \eqref{rels}. 
We note that if  \eqref{lem1}  is not satisfied then the bulk vertex disappears. In the boundary description it exactly corresponds to the case where the fusion rules in the corresponding vertex of the pant decomposition are violated. It explains the notion of the classical fusion polynomial introduced above.

\item  In the limit $\tilde s_1 = 0$ the radial vertex coordinates are related as 
\be
\label{lem2}
\eta_1 = \eta_2[\a \rightarrow  2,\b \rightarrow  \a, 3 \rightarrow  1] \;.
\ee
Using the third relation in \eqref{rels} the proof is straightforward.

\end{itemize}

\subsection{Angular separations} 

Using the definition  $p_\phi = g_{\phi\phi} \dot \phi$ and recalling that the angular momenta are motion constants we find for a given geodesic segment the following angle increment
\be
\label{angular}
\Delta \phi = \pm \frac{p_\phi}{\alpha^2}\dps\int_{\rho^{'}}^{\rho^{''}} \frac{d\rho \cos \rho}{\sin^2 \rho (1 - \frac{p_\phi^2}{\alpha^2}\cot^2 \rho)^{1/2}}\;.
\ee
Here, the overall sign depends on that of $\dot \rho$. 
Explicitly, an angle swept by the geodesic line 
characterized by angular parameter  $s=|p_{\phi}|/\alpha$ is  given by
\be
\label{log}
i \alpha \Delta\phi =\ln
\frac{\sqrt{1-s^2 \cot^2\rho^{''}}-i s \sqrt{1+\cot^2 \rho^{''}}}
{\sqrt{1-s^2 \cot^2\rho^{'}}-i s \sqrt{1+\cot^2 \rho^{'}}}\;.
\ee

Let $\psi_1$ and $\psi_2$ be  angular coordinates of the first  $\a - 1-2$ and the second  $\a- \b - 3$ vertices respectively such that 
$0<\psi_1<w_2 <\psi_2 < w_3$ (recall that we set $w_1 = 0$). Consider angular separations of each geodesic segment. According to Fig. \bref{5bulk}  they are given by  
\be
\Delta\phi_1 =  \psi_1\;,
\quad
\Delta\phi_2 = w_2 - \psi_1\;,
\quad
\Delta\phi_3 = w_3 - \psi_2\;,
\quad
\Delta\tilde \phi_1 = \psi_2 - \psi_1\;,
\quad
\Delta\tilde \phi_2 =  0 \;.
\ee
In particular, one finds the following angular equations 
\be
\label{fc}
\Delta\phi_1 + \Delta \phi_{2} = w_2\;,
\ee
\be
\label{sc}
\Delta\phi_1 + \Delta \phi_{3} + \Delta \tilde \phi_1= w_3\;,
\ee
where each angular separation is given by \eqref{log}.
The above analysis of the equilibrium equations defines radial coordinates $\cot^2 \rho_1$ and $\cot^2 \rho_2$ in terms of the angular momenta.  According to \eqref{log},  the angular separations are functions of two independent momenta, say $s_1$ and $s_3$, and therefore the above equation system can be solved as $s_{1,3} = s_{1,3}(w_2,w_3| \alpha, \epsilon)$.

For later use let us write all ingredients of the above angular equations 
\be
\ba{l}
\dps
i\alpha \Delta \phi_1=\ln \frac{\sqrt{1-s_1^2 \eta_1}-i s_1 \sqrt{1+\eta_1}}{1-i s_1 }\;,\qquad
i\alpha \Delta \phi_2=\ln
\frac{\sqrt{1-s_2^2 \eta_1}-i s_2 \sqrt{1+\eta_1}}
{1-i s_2 }\;,
\\
\\
\dps
i\alpha \Delta \phi_3=\ln
\frac{\sqrt{1-s_3^2 \eta_2}-i s_3 \sqrt{1+\eta_2}}
{1-i s_3 }\;,
\qquad
i\alpha \Delta \tilde\phi_1=\ln
\frac{\sqrt{1-\tilde s_1^2 \eta_2}-i \tilde s_1 \sqrt{1+\eta_2}}
{\sqrt{1-\tilde s_1^2 \eta_1}-i \tilde s_1 \sqrt{1+\eta_1}}\;,
\ea
\ee
where we used notation \eqref{etabar} and \eqref{etatilde}. 

\vspace{-3mm}

\paragraph{First angular condition.} From the condition \eqref{fc} we find  
\begin{eqnarray}
\label{firsteq}
e^{i\alpha w_2}=
\frac{\big(\sqrt{1-s_1^2\, \eta_1}-i s_1 \,\sqrt{1+\eta_1}\big)\big(\sqrt{1-s_2^2\, \eta_1}-i s_2\, \sqrt{1+\eta_1}\big)}
{(1-i s_1) (1-i s_2) }\;.
\end{eqnarray}
The right-hand-side is obviously a unimodular complex number so that  
real and imaginary parts are not independent. It follows that we can analyze either real or imaginary part of equation \eqref{firsteq} depending on simplicity of the corresponding expressions.

We consider the real part of equation \eqref{firsteq}. Denoting  $A=Re[e^{i\alpha w_2}(1-i s_1) (1-i s_2)]$, where 
\be
\label{A}
A=(1-s_1s_2)\cos \alpha w_2+(s_1+s_2) \sin \alpha w_2\;,
\ee
we find out the following irrational equation
\be
\sqrt{1-s_1^2\, \eta_1}\sqrt{1-s_1^2\, \eta_1}  - s_1s_2(1+\eta_1) - A = 0\;.
\ee
This is a typical equation arising from the equilibrium and angular conditions discussed  earlier. Squaring twice one gets rid of the radicals so that the  resulting polynomial equation is linear and its solution reads    
\be
\label{cott}
\eta_1 =\frac{1-(A+s_1 s_2)^2}{s_1^2 +s_2^2 +2s_1s_2(A+s_1s_2)}\;.
\ee
This is to be compared to \eqref{cottilde}. In this way we find our first condition on the angular parameters $s_1$ and $s_2$, namely
\begin{align}
\label{cond1}
\frac{1 - \sigma^2}{s_1^2+s_2^2 -2 s_1s_2 \sigma }=
\frac{1-(A+s_1 s_2)^2}{s_1^2 +s_2^2 +2s_1s_2(A+s_1s_2)}\;.
\end{align}
The most convenient way to analyze the above equation is to introduce variables $u = s_1+s_2$ and $v = s_1 s_2$. The resulting equation is cubic both in $u$ and $v$. It has three real roots and the simplest one is given by
\be
\label{512}
s_2  = \frac{\sigma +  \cos \alpha w_2 +  s_1 \sin \alpha w_2}{-s_1 + s_1 \cos \alpha w_2 - \sin \alpha w_2}\;.
\ee
It can be obtained by equating  $A+v = - \sigma$. 
Two other branches contain non-trivial radicals and should be discarded as they generally  violate the property of $s_{1,2}$ having fixed sign. It is worth noting that the zeros of the denominator in \eqref{512} are given by $s_1 = -\cot (\alpha w_2/2)$.

\vspace{-3mm}

\paragraph{Second angular condition.} Equation \eqref{sc} can be cast into the form 
\be
\label{secondeq}
e^{i\alpha w_3}= \frac{\big(\sqrt{1-s_3^2 \eta_2 }-i s_3 \sqrt{1+\eta_2}\big)
\big(\sqrt{1-\tilde s_1^2 \eta_2}-i \tilde s_1 \sqrt{1+\eta_2}\big)\big(\sqrt{1-s_1^2 \eta_1}-i s_1\sqrt{1+\eta_1}\big)}{(1-i s_3)\big(\sqrt{1-\tilde s_1^2 \eta_1}-i \tilde s_1\sqrt{1+\eta_1}\big)(1-i s_1)}\;.
\ee
As in the previous case we take its real part and find the following relation
\be
\label{cot2}
\eta_2=\frac{1-(s_3 \tilde s_1+B)^2}{s_3^2+\tilde s_1^2 + 2 s_3 \tilde s_1(B+s_3\tilde s_1)}\;,
\ee
where
\be
\label{B}
B=Re\bigg(e^{i\alpha w_3}(1-i s_1)(1-i s_3)\frac{\sqrt{1-\tilde s_1^2 \eta_1}  - i\tilde s_1 \sqrt{1+\eta_1}}
{\sqrt{1-s_1^2 \eta_1}  - is_1 \sqrt{1+\eta_1}}\bigg)\;,
\ee
where $\eta_1 $ is given by \eqref{cott}. This is to be equated to \eqref{1tan}. In this way we obtain our  second angular equation  
\be
\label{seqaneq}
\frac{1-(s_3 \tilde s_1+B)^2}{s_3^2+\tilde s_1^2 + 2 s_3 \tilde s_1(B+s_3\tilde s_1)} = \frac{\tau^2}{s_3^2}\;.
\ee
This equation completely defines coordinate dependence of $s_1$. Indeed, there are two independent momenta chosen to be $s_1$ and $s_3$, while others are related to them through linear conditions \eqref{rels}. The first angular equation relates $s_1$ and $s_2$ by virtue of \eqref{512}. Therefore, the second angular equation fixes $s_1 = s_1(w_2,w_3)$.

\section{The perturbation theory}
\label{sec:deformation}

Our goal is to find solutions to the  second angular equation \eqref{seqaneq}. One possibility to solve this equation is to get rid of all radicals. The resulting equation on $s_2$ is a higher order polynomial equation and  it is  unlikely to be solved exactly. We propose to use a perturbation procedure that helps to find solutions to \eqref{seqaneq}. We consider the \textit{five-line} configuration as a deformation of the \textit{three-line} configuration corresponding to the $4$-point conformal block. Below we recall the 4-pt case \cite{Hijano:2015rla}.  

\subsection{Three-line configuration }
\label{sec:three}

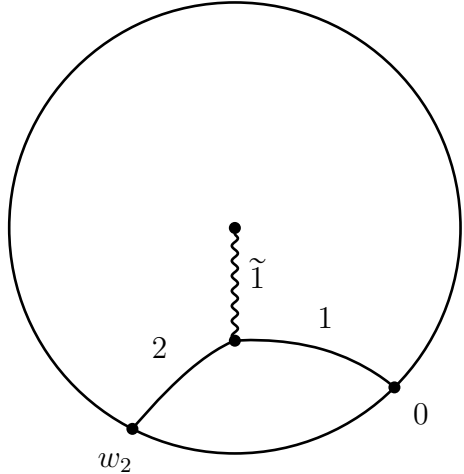
\begin{figure}[H]
\centering
\begin{tikzpicture}[line width=1pt]
\draw (0,0) circle (3cm);

\foreach \a in {1,2,...,40}{
\draw (\a*360/40: 3.5cm) coordinate(N\a){};
\draw (\a*360/40: 3cm) coordinate(K\a){};
\draw (\a*360/40: 2cm) coordinate(L\a){};
\draw (\a*360/40: 1cm) coordinate(M\a){};
\draw (\a*360/40: 1.5cm) coordinate(I\a){};}


\draw plot [smooth, tension=1.0, line width=1pt] coordinates {(K35) (L34) (I30)};
\draw plot [smooth, tension=1.0, line width=1pt] coordinates {(K27) (L28) (I30)};


\draw [smooth, tension=1.0, line width=1pt, decorate, decoration = {snake, segment length = 2mm, amplitude=0.4mm}] (I30)  -- (0,0);


\fill (I30) circle (0.8mm);

\fill (K35) circle (0.8mm);
\fill (K27) circle (0.8mm);
\fill (0,0) circle (0.8mm);

\draw (1.2,-1.2) node {$1$};
\draw (-1.0,-1.6) node {$2$};
\draw (0.3,-0.6) node {$\a$};

\draw (N35) node{$0$};
\draw (N27) node{$w_2$};

\end{tikzpicture}
\caption{Three-line graph. Solid lines represent external particles, wavy lines represent intermediate particles, \cite{Hijano:2015rla}.} 
\label{3bulk}
\end{figure}

In this case there are  two light external fields with dimensions $\epsilon_1$ and $\epsilon_2$ and one intermediate light field with dimension $\tilde\epsilon_1$. The respective graph is depicted on Fig. \bref{3bulk}. 

\noindent Here, the equilibrium and the angular equations read 
\be
\label{3verteq}
\epsilon_1 \sqrt{1-s_1^2\, \eta} + \epsilon_2 \sqrt{1-s_2^2 \, \eta} = \tilde\epsilon_1\;,
\qquad
\epsilon_1 s_1  - \epsilon_2 s_2 = 0\;,
\ee
\be
\label{3angeq}
e^{i\alpha w_2}=
\frac{(\sqrt{1-s_1^2\, \eta }-i s_1 \sqrt{1+\eta})(\sqrt{1-s_2^2\, \eta }-i s_2 \sqrt{1+ \eta})}
{(1-i s_1)(1-i s_2)}\;,
\ee
where the radial coordinate of the vertex is 
\be
\eta =  - \frac{\Pi(\epsilon_1, \epsilon_2, \tilde\epsilon_1)}{4\, \tilde\epsilon_1^2 \,\epsilon_1^2 \,s_1^2}\;,
\ee
and $\Pi(\epsilon_1, \epsilon_2, \tilde\epsilon_1)$ is the fusion polynomial \eqref{fusion}. We note that the radial coordinate is  
\be
\label{lem3}
\eta = \eta_2 [3 \rightarrow 1, \a \rightarrow 2, \b \rightarrow \a]\;.
\ee

In order to simplify the analysis  we consider  equal external dimensions. Then, the solution to the above equations is 
\be
\label{3cases1}
\epsilon_1 = \epsilon_2\,:
\qquad
\quad
\dps s_1 = s_2 = - \cot \theta_2  + \frac{\tilde\epsilon_1}{2\epsilon_1 \sin \theta_2}\;,
\qquad
\theta_2  \equiv \frac {\alpha w_2}{2}\;.
\ee 
The total action in this case is $S_0 = 2\epsilon_1 S_1 + \tilde\epsilon_1 S_{\a}$. Using  \eqref{lambda} and \eqref{3cases1} we find 
\be
\label{56}
S_1 =   - \ln \sin \theta_2  + \ln \sqrt{1 - \frac{\tilde\epsilon^2_1}{4\epsilon_1^2}}- \ln 2\Lambda \;,
\qquad
S_{\a} = \ln\tan \frac{\theta_2}{2} + \ln \sqrt{\frac{\epsilon_1 + \tilde\epsilon_1/2}{\epsilon_1 -\tilde\epsilon_1/2}} \;,
\ee
where $\Lambda \rightarrow \infty$ is the boundary regulator. We observe that all conformal dimensions arise as additive contributions and  can therefore be neglected. This is why coordinate dependent terms in \eqref{56} coincide with those in \eqref{rad} and \eqref{arc}.  
Modulo irrelevant coordinate independent terms the total action is  given  by  
\be
\label{S0}
S_0(w_2) = - 2\epsilon_1 \ln \sin \theta_2 + \tilde\epsilon_1 \ln\tan \frac{\theta_2}{2}\;.
\ee

\subsection{Truncation to the 4-pt case}
\label{sec:truncation}

We choose one of the external fields (which is in the middle of the pant decomposition) to be an identity operator, while  intermediate fields of the corresponding vertex get equal dimensions. The vertex disappears while  $n$-pt block  goes to $(n-1)$-pt block. If the identity external field is chosen to be on the edge of the block  then  one should equate dimensions of one intermediate and one external field of the corresponding vertex.

Let us see how it works in the bulk analysis. We consider the vertex  equations and set conformal dimensions of various external fields to zero. The angular equations get corresponding modification. There are three types of truncation.  

\vspace{-3mm}

\paragraph{The $\epsilon_3 = 0$ case.} This configuration corresponds to 4-pt conformal block provided that external dimensions are equal to each other,  $\tilde\epsilon_1 = \tilde\epsilon_2$.  
The vertex equations \eqref{rels}, \eqref{vertcoord11} and \eqref{vertcoord21} can be written as 
\be
\label{chet}
\ba{c}
\dps
\epsilon_3\sqrt{1 - s_3^2 \eta_2} +\tilde\epsilon_1\sqrt{1-\tilde s_1^2 \eta_2} = \tilde\epsilon_1\;,
\qquad
\epsilon_3 s_3  - \tilde\epsilon_1 \tilde s_1 = 0\;,

\ea
\ee
and
\be
\label{subb}
\ba{c}
\epsilon_1 \sqrt{1-s_1^2 \eta_1} + \epsilon_2 \sqrt{1-s_2^2 \eta_1} = \tilde\epsilon_1 \sqrt{1-\tilde s_1^2 \eta_1}\;,
\qquad
\epsilon_1 s_1  - \epsilon_2 s_2 -\tilde\epsilon_1 \tilde s_1 = 0\;,
\ea
\ee
Setting   $\epsilon_3 = 0$ we reproduce the 4-pt case \cite{Hijano:2015rla}. Equations \eqref{chet} are satisfied identically so that the vertex $\a - \b - 3$ disappears. In this limit the intermediate line $\a$ becomes radial, \textit{i.e.}, $\tilde s_1=0$, so that using \eqref{lem2} and \eqref{lem3}, equations \eqref{subb} are reduced to  \eqref{3verteq}.

\vspace{-3mm}

\paragraph{The $\epsilon_2 = 0$ case.} This configuration corresponds to the 4-pt conformal block  with $\tilde\epsilon_1 = \epsilon_1$. In this case, the  vertex equations can be cast into the form 
\be
\ba{c}
\dps
\epsilon_3 \sqrt{1 - s_3^2 \eta_2} +\epsilon_1\sqrt{1-\tilde s_1^2 \eta_2} = \tilde\epsilon_2\;,
\qquad
\epsilon_3 s_3  -  \epsilon_1 \tilde s_1 = 0\;, 
\ea
\ee
and
\be
\ba{c}
\epsilon_1\sqrt{1-s_1^2 \eta_1} + \epsilon_2 \sqrt{1-s_2^2 \eta_1} = \epsilon_1\sqrt{1-\tilde s_1^2 \eta_1}\;,
\qquad
\epsilon_1 s_1  - \epsilon_2  s_2 -\epsilon_1 \tilde s_1 = 0\;. 

\ea
\ee
For $\epsilon_2 = 0$ the second pair of equations trivializes, while the first one goes to that of the 4-pt case provided $\tilde s_1 = 0$. 

\vspace{-3mm}

\paragraph{The $\epsilon_1 = 0$ case.} This configuration does not correspond to 4-pt conformal block because 
the equilibrium  equation $\tilde\epsilon_1 \tilde s_1 + \epsilon_2 s_2 = 0$ has no admissible solutions:   all $s_I \geq 0$ so that the only solution here is $\tilde s_1 = s_2 = 0$ that corresponds to  merging of the lines $\a$ and $2$ into single radial line. Therefore, the configuration depicted on Fig. \bref{3bulk} can not be  reproduced. To have  a correct graph  of the 4-pt conformal block we need to modify the initial configuration by changing the slope of line  2.

\subsection{Five-line configuration as a deformation}

A five-line configuration can be considered as a deformation of the  three-line configuration with respect to one of the external conformal dimensions. In what  follows, we explicitly consider the case where $\epsilon_3$ is the deformation parameter and other conformal dimensions are
\be
\tilde\epsilon_1 = \tilde\epsilon_2\;,
\qquad
\epsilon_1 = \epsilon_2\;.
\ee
The first condition here is required for consistency of the truncation. The second condition is imposed to simplify our consideration. \footnote{The same constraints have been used in the boundary computations \eqref{PPPP}.} Since two heavy operators produce the background and other operators are light, the operator associated to line $3$ should be considered as superlight. In other words, the  true deformation parameter is 
\be
\nu = \frac{\epsilon_3}{\tilde\epsilon_1}\;.
\ee

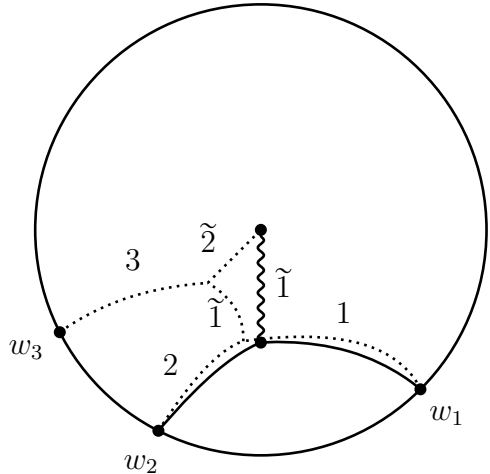
\begin{figure}[H]
\centering
\begin{tikzpicture}[line width=1pt]
\draw (0,0) circle (3cm);

\foreach \a in {1,2,...,40}{
\draw (\a*360/40: 3.5cm) coordinate(N\a){};
\draw (\a*360/40+5: 2cm) coordinate(A\a){};
\draw (\a*360/40: 3cm) coordinate(K\a){};
\draw (\a*360/40: 2cm) coordinate(L\a){};
\draw (\a*360/40: 1.5cm) coordinate(I\a){};
\draw (\a*360/40: 1.1cm) coordinate(J\a){};
\draw (\a*360/40: 1cm) coordinate(M\a){};
;}


\draw plot [smooth, tension=1.0, line width=1pt] coordinates {(K35) (L34) (I30)};
\draw plot [smooth, tension=1.0, line width=1pt] coordinates {(K27) (L28) (I30)};

\draw [dotted] plot [smooth, tension=1.0, line width=1pt] coordinates {(K35) (A34) (I29)};
\draw [dotted] plot [smooth, tension=1.0, line width=1pt] coordinates {(K27) (A27) (I29)};

\draw [dotted] plot [smooth, tension=1.0, line width=1pt,decoration = {snake, segment length = 2mm, amplitude=0.4mm}] coordinates {(K23) (L23) (M25)};

\draw [smooth, tension=1.0, line width=1pt, decorate, dotted] (M25)  -- (0,0);

\draw [dotted] plot [smooth, tension=1.0, line width=1pt] coordinates {(I29)(J28) (M25)};


\draw  [smooth, tension=1.0, line width=1pt, decorate, decoration = {snake, segment length = 2mm, amplitude=0.4mm}]  (I30) -- (0,0);


\fill (I30) circle (0.8mm);

\fill (K35) circle (0.8mm);
\fill (K27) circle (0.8mm);
\fill (K23) circle (0.8mm);
\fill (0,0) circle (0.8mm);

\draw (1.1,-1.1) node {$1$};
\draw (-1.2,-1.8) node {$2$};
\draw (-1.7,-0.4) node {$3$};
\draw (-0.7,-0.1) node {$\b$};
\draw (-0.6,-1.1) node {$\a$};
\draw (0.3,-0.7) node {$\a$};
\draw (N23) node {$w_3$};
\draw (N27) node {$w_2$};
\draw (N35) node {$w_1$};

\end{tikzpicture}
\caption{A deformation method. Vertex $\a-\b-3$ originates from the seed vertex point attached to the radial line $\a$. The deformation produces lines $\a$ and $\b$ from the original line $\a$ by pulling  the seed vertex point using  line $3$. Solid lines correspond to the $4$-pt case, while dotted ones indicate the $5$-pt deformation.} 
\label{35bulk}
\end{figure}

The deformation of the three-line configuration depicted on Fig. \bref{3bulk} can be visualized as the  "seed" vertex attached to the radial line $\a$. Pulling it out splits the radial line $\a$ into radial line $\b$ and curved  line $\a$, and produces external line $3$ as depicted in  Fig. \bref{35bulk}. The lines of the resulting five-line configuration are characterized  by the deformed angular momenta 
\be
\label{expan}
s_I = b_I + \nu c_I + \cO(\nu^2)\;,
\qquad I = 1,2,3,\a,\b\;,
\ee
where $b_I$ are the angular  momenta of the seed three-line configuration and $c_I$ are corrections. Note that $\tilde s_2=b_{\b} =0$ remain intact, and the seed line $\a$ is radial so that $b_{\a} = 0$. By convention, $b_3$ is the seed momentum  assigned to line $3$. The total action reads
\be
S(w_2, w_3)= S_0(w_2) + \nu S_1(w_2,w_3) + \cO(\nu^2)\;, 
\ee
where $S_0 = S_0(w_2)$ is the action of the three-line  configuration, while $S_1(w_2,w_3)$ is a correction. Note that a position of the superlight field $w_3$ enters the action through the correction only.

The idea behind the perturbation method is to replace finding solutions to higher order algebraic equations by solving linear recurrence equations imposed on the corrections.  
Using the approximation \eqref{expan} we expand the angular equations up to linear terms in $\nu$ and find out that the seed solution satisfies the original algebraic equation for $\nu =0$  while first order terms are linear equations expressing the corrections through the seed solution. Higher-order corrections are also subjected to linear recurrence equations and can be directly found. Therefore, the most complicated part of the problem is to find a  seed solution which in our case is explicitly known and corresponds to the 4-pt configuration described in Sect. \bref{sec:three}.

\section{Perturbative solution}
\label{sec:pert}

Now we are going to identify the position of the seed vertex attached to the radial line $\a$, see Fig. \bref{35bulk}. To this end, we  consider $s_1$ and $s_3$ as independent angular momenta. From  the equilibrium equations \eqref{chet} and \eqref{subb}  we find angular positions of the  vertices 
\be
\label{etas}
\eta_1 =  \frac{1 - \varkappa^2/4}{s_1^2 + \nu^2 s_3^2  - \nu \varkappa s_1 s_3}\;,
\qquad\qquad 
\eta_2  = \frac{1 - \nu^2/4}{s_3^2}\;,
\ee
where $\varkappa = \tilde\epsilon_1/\epsilon_1$, cf. \eqref{solution} and \eqref{1tan}.
We see that the the second vertex position can be smoothly continued  to $\nu=0$ to obtain the seed vertex $\eta_2 = 1/s_3^2$. 

The seed vertex position can be used to find the corresponding seed angular momenta values. Consider the angular equations \eqref{firsteq} and \eqref{secondeq} for $\nu=0$. They can be cast into the form
\begin{eqnarray}
\label{firsteq0}
e^{i\alpha w_2/2}=
\frac{\sqrt{1-s_1^2 \eta_1}-i s_1 \sqrt{1+\eta_1}}
{1-i s_1}\;,
\end{eqnarray}
\begin{eqnarray}
\label{secondeq0}
e^{i\alpha (w_3 - w_2/2)}=
\frac{\sqrt{1-s_3^2 \eta_2}-i s_3 \sqrt{1+\eta_2}}
{1-i s_3}\;.
\end{eqnarray}
Equation \eqref{firsteq0} reproduces the angular equation of the three-line case \eqref{3angeq} provided $s_1= s_2$. Equation \eqref{secondeq0} is not seen in the three-line case and appears because our angular equations  do not follow from the original action contrary to the equilibrium equations. It follows that for $\nu = 0$ there remains  a residual angular equation \eqref{secondeq0} that helps to define the seed momentum $s_3$. From \eqref{secondeq0} we find   
\be
\label{ss3}
s_3 =  - \cot(2\theta_3 - \theta_2)\;, \qquad \epsilon_3 = 0\;,
\ee
where we introduced notation 
\be
\label{thet}
\theta_k = \frac{\alpha w_k}{2}\;,
\qquad
k = 1,2,3\;.
\ee
Note that the seed momentum $s_3$ is not vanishing despite that $\epsilon_3 = 0$. Geometrically,  it means that we can add such a line to the three-line graph but its weight in the total action is kept zero. The deformation makes $\epsilon_3$ non-vanishing so that the line starts to be seen.

Finally,  using the three-line solution \eqref{3cases1} and notation \eqref{expan} we write down  all the seed angular momenta as 
\be
\label{sss}
b_1 = b_2 = - \cot \theta_2 +  \frac{\varkappa}{ 2\sin \theta_2}\;, 
\qquad
b_3 =  - \cot(2\theta_3 - \theta_2)\;, 
\qquad
b_{\a} = b_{\b} =0\;.
\ee

In what follows we consider the angular equations for $\nu\neq 0$ and apply the perturbation expansion around the seed solution \eqref{sss}.  

\vspace{-3mm}

\paragraph{Expansion of the  angular equations.} Using notation \eqref{thet} the real part of the angular equation \eqref{firsteq} for $\nu \neq 0$ is represented as 
\be
\label{eqeq}
\sqrt{1-s_1^2 \,\eta_1}\,\sqrt{1-s_2^2\, \eta_1} - s_1 s_2 (1+\eta_1)  = (1-s_1 s_2) \cos2\theta_2 +(s_1+s_2)\sin 2\theta_2\;.
\ee 
Substituting \eqref{etas}, \eqref{sss}, and \eqref{expan} into equation \eqref{eqeq} and keeping first-order terms in the deformation parameter $\nu$ we find that the original equation is reduced to the following linear relation 
\be
\label{512light}
2 c_1 - \varkappa b_3 = 0\;.
\ee
As expected, the same relation follows from the exact solution \eqref{512} which in the case under consideration takes the form 
\be
s_2 = - \cot \theta_2 \bigg(1 - \half\, \frac{\varkappa^2}{\sin 2 \theta_2\,(s_1 + \cot \theta_2) }\bigg)\;.
\ee

Now, it is convenient to identically represent the second angular equation \eqref{secondeq} as 
\be
\ba{l}
e^{2i\theta_3} (1-is_1)(1-is_3)(\sqrt{1-\nu^2 s_3^2 \eta_1} - i \nu s_3\sqrt{1+\eta_1}) = 
\\
\\
\hspace{2cm} =(\sqrt{1-s_3^2 \eta_2} - is_3 \sqrt{1+\eta_2})(\sqrt{1-\nu^2 s_3^2 \eta_2} - i \nu s_3\sqrt{1+\eta_2})(\sqrt{1-s_1^2 \eta_1} - is_1 \sqrt{1+\eta_1})\;.
\ea
\ee
Taking  \eqref{512light} into account  we find that  in the first order  approximation  the real part of the above equation  is a linear equation on $c_3$ solved as 
\be
c_3 =  \frac{3+\cos(2\theta_2 - 4\theta_3) - 2 \cos(2\theta_2 - 2\theta_3) - 2\cos(2\theta_3)}{4\sin^3(\theta_2-2\theta_3)}\;.
\ee

\vspace{-3mm}

\paragraph{The deformed angular parameters.} The first order solution to the 5-pt configuration reads
\be
s_1 = - \cot \theta_2 +  \frac{\varkappa}{ 2\sin \theta_2} - \frac{\nu \varkappa}{2} \cot(2\theta_3- \theta_2) + \cO(\nu^2)\;,
\ee

\be
s_3 = - \cot(2\theta_3 - \theta_2) + \nu \frac{\cos(2\theta_2 - 4\theta_3) - 2 \cos(2\theta_2 - 2\theta_3) - 2\cos2\theta_3+3}{4\sin^3(\theta_2-2\theta_3)}+ \cO(\nu^2)\;,
\ee
and
\be
s_2 = s_1 - \nu\varkappa s_3\;,
\qquad
\tilde s_1 = \nu s_3\;,
\qquad
\tilde s_2 = 0\;.
\ee

\vspace{-3mm}

\paragraph{The action.} Finally, using representation \eqref{lambda} we explicitly write down  the total action \eqref{act5} as 
\be
\label{actionFULL1}
S(w_2, w_3) =  \epsilon_1 S_1(w_2, w_3) + \epsilon_1 S_2(w_2, w_3)+ \epsilon_3 S_3(w_2, w_3) +\tilde\epsilon_1 S_{\a}(w_2, w_3)+ \tilde\epsilon_1 S_{\b}(w_2, w_3) \;,
\ee
where    
\be
\label{L1}
S_1 =  - \ln \frac{\sqrt{\eta_1}}{\sqrt{1+\eta_1}+\sqrt{1  - s_1^2\eta_1}} - \ln 2\Lambda\;,
\ee

\vspace{5mm}

\be
\label{L2}
S_2 =  - \ln \frac{\sqrt{\eta_1}}{\sqrt{1+\eta_1}+\sqrt{1- s_2^2\eta_1}} - \ln 2\Lambda \;,
\ee
\vspace{5mm}
\be
\label{L3}
S_{3} = - \ln \frac{\sqrt{\eta_2}}{\sqrt{1+\eta_2}+\sqrt{1- s_3^2\eta_2}} - \ln 2\Lambda\;,
\ee

\vspace{5mm}

\be
\label{La}
S_{\a} = \ln \frac{\sqrt{\eta_1}}{\sqrt{1+\eta_1}+\sqrt{1- \nu^2 s_3^2\eta_1}}- \ln \frac{\sqrt{\eta_2}}{\sqrt{1+\eta_2}+\sqrt{1- \nu^2 s_3^2\eta_2}} \;,
\ee

\vspace{5mm}

\be
\label{Lb}
S_{\b} = \ln \frac{\sqrt{\eta_2}}{1+ \sqrt{1+\eta_2}}\;.
\ee
Here, $\Lambda$ is the cutoff parameter, $\Lambda \rightarrow \infty$. 
For $\nu = 0$ we  note that  $S_{\a}+S_{\b} = \ln \frac{\sqrt{\eta_1}}{1+ \sqrt{1+\eta_1}}$ and this formula gives the length of the radial line \eqref{rad}. 
Also, for $\nu=0$ the action $S_3$ does not contribute to the total action, while $S_1 = S_2$ and $2S_1$ gives the length of the arc, cf. \eqref{arc}. We conclude that for  $\nu=0$ the action \eqref{actionFULL1} is identified with the 4-pt case action \eqref{S0}.

In the first order in $\nu$ the above actions are given by
\be
S_1 = -\ln\sin \theta_2  + \nu \cot(\theta_2 - 2 \theta_3) + \ln \sqrt{1 - \frac{\tilde\epsilon^2_1}{4\epsilon_1^2}}- \ln 2\Lambda+ \cO(\nu^2)\;,
\ee

\be
S_2 = -\ln\sin \theta_2  - \nu \cot(\theta_2 - 2 \theta_3) + \ln \sqrt{1 - \frac{\tilde\epsilon^2_1}{4\epsilon_1^2}}- \ln 2\Lambda+ \cO(\nu^2)\;,
\ee

\be
S_{3} = -\ln \sin(2 \theta_3-\theta_2)- \ln 2\Lambda+\cO(\nu)\;,
\ee

\be
\ba{l}
\dps
S_{\a} = \ln\tan\frac{\theta_2}{2} -\ln\tan(2 \theta_3-\theta_2) + \ln \sqrt{\frac{\epsilon_1 + \tilde\epsilon_1/2}{\epsilon_1 -\tilde\epsilon_1/2}}
\\
\\
\dps
\hspace{1cm}+\,\nu\, \frac{ \cos(2\theta_2 - 4\theta_3) - 2 \cos(2\theta_2 - 2\theta_3) - 2\cos2\theta_3+3}{4 \sin^2(\theta_2 - 2\theta_3)} + \cO(\nu^2)\;,
\ea
\ee

\be
\ba{l}
\dps
S_{\b} = \ln\tan(2 \theta_3-\theta_2)  
\\
\\
\dps
\hspace{1cm} - \,\nu\, \frac{ \cos(2\theta_2 - 4\theta_3) - 2 \cos(2\theta_2 - 2\theta_3) - 2\cos2\theta_3+3}{4 \sin^2(\theta_2 - 2\theta_3)} + \cO(\nu^2)\;.
\ea
\ee
Note that in the first order the actions $S_{1,2}$  and $S_{\a,\b}$  get corrections, while their sums $S_1 +S_2$ and $S_{\a}+S_{\b}$ remain intact. The regulator $\Lambda$ appears for the external lines only. 

Modulo coordinate independent terms we arrive at the following total action 
\be
\label{finact}
S(w_2, w_3) =  - 2 \epsilon_1 \ln\sin \theta_2 + \tilde\epsilon_1  \ln\tan\frac{\theta_2}{2}
- \epsilon_3 \ln \sin(2 \theta_3-\theta_2) + \cO(\nu^2)\;. 
\ee  
The zeroth order part obviously coincides with the three-line action \eqref{S0}. It is parameterized by the variable $\theta_2$, while the first order  correction depends on coordinates $w_{2,3}$ through the combination $2\theta_3 - \theta_2$ only.   
Note that the resulting action \eqref{finact} has no pole $1/\tilde\epsilon_1$ while it appears in higher order terms, cf. \eqref{5ptdec1}.

According to the general prescription \eqref{block-action}, the action \eqref{finact} is related  to the conformal block as
\be
f_{\delta}(q_1,q_2)\sim  - S(\theta_1,\theta_2)\;,
\ee
where the first terms of the conformal block on the plane are given in  \eqref{finblock1} and \eqref{finblock2}. The identification is achieved (up to irrelevant coordinate-independent constants which can be absorbed in the integration constants and also taking into account standard conformal block prefactor which is not assumed to exponentiate)
 by the following conformal transformations to the plane
\be
\theta_2=\frac{i\alpha}{2}\ln (1-q_1 q_2)\;, \qquad \theta_3=\frac{i\alpha}{2}\ln(1-q_2)\;.
\ee
The perturbation procedure which allows to evaluate the bulk configuration \eqref{finact} together with its boundary counterpart 
computation of the dual classical conformal block \eqref{finblock1} and \eqref{finblock2} represent the main result of this paper. In the next section we discuss its possible generalization to the $n$-point case.

\section{Towards multi-line configurations}
\label{sec:multi}

The $n=5$ analysis in the previous sections shows the way it generalizes to the $n$-point case. According the graph in Fig. \bref{bulk} we discuss a multi-line configuration with $n-2$ external legs using the following notation and conventions.  

\begin{itemize}

\item There are $n-2$ external lines with index $i = 1, 2, ... , n-2$ provided that lines $1$ and $2$ are on the rightmost edge of the graph. There are $n-3$ intermediate lines with index  $\tilde{1}, \2, ... , \tilde{n-3}$ provided that $(\tilde{n-3})$-th line is identified with the radial line ending in the singularity. Both external and intermediate lines are enumerated by the collective index $I = 1,...,n-2, \1,...,\tilde{n-3}$.

\item There are $n-3$ vertices with radial positions denoted $\eta_{i} = \cot^2 \rho_{i}$, where $i = 1,..., n-3$.

\item $i$-th  vertex with $i = 2,..., n-4$ has three incoming lines: one external line $i+1$, two intermediate lines $\tilde{i-1}$ and $\tilde{i+1}$. The $i=1$ vertex has three incoming lines: two external lines $1$ and $2$, one intermediate line $\1$. The $i=n-3$ vertex has three incoming lines: one external line $n-2$, two intermediate lines $\tilde{n-4}$ and  $\tilde{n-3}$.  

\end{itemize}

\noindent Consider first the equilibrium conditions for any three fields
\be
\label{eqcondi}
\epsilon_K\sqrt{1-s_K^2 \eta} + \epsilon_I\sqrt{1-s_I^2 \eta} = \epsilon_J\sqrt{1-s_J^2  \eta}\;,
\ee
\be
\label{gens}
\epsilon_I s_I +\epsilon_J s_J - \epsilon_K s_K =0 \;,
\ee
where $I\neq J\neq K$. Using the fusion polynomials \eqref{fusion} the general solution to equations \eqref{eqcondi} and \eqref{gens} is given by   
\be
\label{etai}
\eta  =\frac{\Pi(\epsilon_K, \epsilon_I, \epsilon_J)}{4\epsilon_I \epsilon_K\big( s_I s_K(\epsilon_I^2 +\epsilon_K^2 - \epsilon_J^2) -\epsilon_I \epsilon_K (s_I^2 +s_K^2)\big)}\;,
\ee
cf. \eqref{cottilde}. As there are vertices of three different types, the particular values of indices should be properly taken into account. It follows that there are two outmost positions $\eta_1$ with $I=2, J=\1, K=1$ and  $\eta_{n-3}$ with $I = n-2, J = \tilde{n-3}, K = \tilde{n-4}$, along with intermediate positions $\eta_{i}$,  $i=2,..., n-4$, with $I = i+1, J =\tilde i, K = \tilde{i-1}$.    

Explicit form of equations \eqref{gens} is given by 
%
%
%
%
%
%

\be
\label{linrelss}
\ba{l}
\tilde s_{{n-3}} = 0\;, 
\qquad
\epsilon_{2} s_{2} +\tilde \epsilon_{{1}}\tilde s_{{1}} - \epsilon_{1} s_{1} = 0\;,
\\
\\
\epsilon_{i} s_{i} +\tilde \epsilon_{i-1}\tilde s_{i-1} - \tilde \epsilon_{i-2}\tilde s_{i-2} = 0\;,
\qquad
i =3, ..., n-2\;.

\ea
\ee

Now, consider the angular separations of each geodesic segment. Let $\Delta \phi_i$ be an angular separation of the $i$-th external line and $\Delta \tilde \phi_{i}$ be an angular separation of the intermediate line $\tilde i$. Introducing angular positions of the vertices $\psi_{i}$ with $i = 1,...,n-3$ we  define ($w_1 =0$)   
\be
\Delta\phi_1 = \psi_1\;,
\qquad
\Delta \phi_i  = w_i - \psi_{i-1}\;, 
\qquad
i = 2, ..., n-2\;,
\ee
\be
\Delta\tilde \phi_{{n-3}} = 0\;,
\qquad
\Delta\tilde \phi_{i} = \psi_{{i+1}} - \psi_{i}\;,
\qquad
i = 1,..., n-4\;.
\ee
In particular, assuming that $\psi_{i}< w_{i+1}$ one finds the angular equation system
\be
\label{angeqGEN}
\Delta \phi_i + \Delta \tilde \phi_{{i-2}} + \Delta \tilde\phi_{{i-3}} + ... + \Delta \tilde \phi_{1}+ \Delta \phi_1 = w_{i}\;, 
\qquad
i = 2, ..., n-2\;.
\ee
In total, there are $n-3$ angular equations that exactly matches the number of boundary attachments minus one. Depending on the particular vertex there are three types of the angular separations entering equation \eqref{angeqGEN}. Using \eqref{log} we find   
\be
\ba{l}
\dps
i\alpha \Delta \phi_1=\ln \frac{\sqrt{1-s_1^2 \eta_1}-i s_1 \sqrt{1+\eta_1}}{1-i s_1 }\;,\\
\\
\dps
i\alpha \Delta \phi_k=\ln \frac{\sqrt{1-s_k^2 \eta_{k-1}}-i s_k \sqrt{1+\eta_{k-1}}}{1-i s_k }\;,
\qquad k  = 2,..., n-2\;,\\
\\
\dps
i\alpha \Delta \tilde\phi_{k}=\ln\frac{\sqrt{1-\tilde s_{k}^2\, \eta_{k+1}}-i \tilde s_{k} \,\sqrt{1+ \eta_{k+1}}}
{\sqrt{1-\tilde s_{k}^2\, \eta_{k}}-i \tilde s_{k} \,\sqrt{1+\eta_{k}}}\;,
\qquad k  = 2,..., n-3\;.\\
\ea
\ee
From the above relations we compose the following system of $n-3$ irrational equations  
\be
\label{angeqFant}
e^{2i\theta_k} = \frac{\sqrt{1-s_1^2 \eta_1}-i s_1 \sqrt{1+\eta_1}}{1-i s_1 }
\frac{\sqrt{1-s_k^2 \eta_{k-1}}-i s_k \sqrt{1+\eta_{k-1}}}{1-i s_k }
\prod_{m=1}^{k-2}\frac{\sqrt{1-\tilde s_{m}^2\, \eta_{m+1}}-i \tilde s_{m} \sqrt{1+ \eta_{m+1}}}
{\sqrt{1-\tilde s_{m}^2\, \eta_{m}}-i \tilde s_{m} \sqrt{1+\eta_{m}}}\;,
\ee
where $k = 2,...,n-2$, and  $\theta_k = \alpha w_k/2$\;.

The solution to the angular equation system \eqref{angeqFant} is given by $n-3$ momenta, say, external parameters $s_i$ for $i = 1,2,...,n-3$, expressed as functions of all boundary attachments 
$s_i = s_i(w_2,...,w_{n-2})$. All other angular parameters are restored via linear relations \eqref{linrelss}. Indeed, there are $n-2$ equations \eqref{linrelss} and  their number is equal to the number of vertices. They fix  $n-2$ angular parameters in terms of others $n-3$ determined by the angular equations. Therefore, all angular momenta are fixed in terms of $n-3$ boundary attachments $w_i$.

Finally, using explicit expressions for all $s_I=s_I(w)$ and $\eta_i = \eta_i(w) $ one finds the weighted length of the multi-line graph 
\be
\label{actFant}
\ba{r}
\dps
S(w_2,...,w_{n-2}) =  - \epsilon_1\ln \frac{ \sqrt{\eta_1}}{\sqrt{1+\eta_1} +  \sqrt{1 - s_1^2 \eta_1}} - \sum_{i=1}^{n-3}\epsilon_{i+1} \ln \frac{ \sqrt{\eta_{i}}}{\sqrt{1+\eta_{i}} +  \sqrt{1 - s_{i+1}^2 \eta_{i}}}
\\
\\
\dps
\hspace{9mm}+\sum_{i=1}^{n-3} \tilde \epsilon_i \Big[\ln \frac{\sqrt{\eta_i}}{\sqrt{1+\eta_i}+\sqrt{1- \tilde s_{i}^2\eta_i}}- \ln \frac{\sqrt{\eta_{i+1}}}{\sqrt{1+\eta_{i+1}}+\sqrt{1- \tilde s_{i}^2\eta_{i+1}}}\Big]\;,
\ea
\ee
where each component is given by formula \eqref{lambda}. Also, we neglected all the regulator terms $\ln 2\Lambda$. Mapping attachment  points  on the cylinder into the corresponding points of the sphere $z_i = z_i(w_k)$ and using projective coordinates $q_i  = q_i(z_j)$, where $i,j,k = 1,...,n-3$ the resulting action is to be identified with the $n$-point heavy-light  conformal block associated to \eqref{FF}  according to the general formula \eqref{block-action}.

It is worth noting that the total action \eqref{actFant} contains the same radicals  as the angular equations \eqref{angeqFant} as well as the equilibrium equation \eqref{eqcondi}. Moreover, the $n=5$ case shows that the expressions for angular momenta and  separate actions can be rather complicated while the total action is simple enough. It suggests that 
solving the angular equations explicitly is not needed because their constituents can be  used directly in the total action.

\section{Conclusion}
\label{sec:conclusion}

In this paper we have analyzed  the AdS/CFT correspondence between $n$-point heavy-light classical conformal blocks on the boundary and classical worldline actions described by a graph with $n$ worldlines embedded in the bulk. We have proposed the general identification between the pant decomposition on the boundary and the corresponding multi-line graph in the bulk. In particular, we have written down the general system of equations describing  the dynamics of probes in the bulk background. On the boundary side, the classical conformal blocks are conveniently analyzed using the AGT combinatorial representation.  

We have performed explicit computations in the $n=5$ case establishing the correspondence in the first order in the conformal dimension of one of fields while keeping other dimensions arbitrary.  
It exactly corresponds to deforming the four-point classical conformal block by adding a superlight external field to yield the five-point classical conformal block. The same perturbative procedure is employed in the bulk where we start with the corresponding four-line worldline configuration while one external line  and one intermediate line are produced in the course of the deformation.

It would be interesting to find an exact solution to $n=5$ equations in the bulk and further compare with the boundary computations beyond the perturbation theory. More generally, one can analyze the $n$-point equation system that we proposed in Sec. \bref{sec:multi} and describe perturbative or exact  solutions.   
Also, it is natural  to study $n$-point heavy-light classical blocks with any number of heavy fields and  elaborate on their bulk interpretation.   

\vspace{5mm}

\noindent \textbf{Acknowledgements.}  The work of K.A. was supported by the Russian Science Foundation grant 14-42-00047 in association with  Lebedev Physical Institute. The work of V.B. was performed at the Institute for Information
Transmission Problems with the financial support of the Russian Science
Foundation (Grant No.14-50-00150). V.B. is grateful to Chaiho Rim for the warm hospitality at the  Sogang University, Seoul and for interesting discussions. 

\providecommand{\href}[2]{#2}\begingroup\raggedright
\addtolength{\baselineskip}{-3pt} \addtolength{\parskip}{-1pt}


\providecommand{\href}[2]{#2}\begingroup\raggedright\endgroup

\end{document}